\documentclass[fleqn,usenatbib]{mnras}

\usepackage{newtxtext,newtxmath}

\usepackage[T1]{fontenc}

\DeclareRobustCommand{\VAN}[3]{#2}
\let\VANthebibliography\thebibliography
\def\thebibliography{\DeclareRobustCommand{\VAN}[3]{##3}\VANthebibliography}

\usepackage{graphicx}	
\usepackage{amsmath}	
\usepackage{xcolor}
\usepackage[normal]{threeparttable}
\usepackage{orcidlink}

\title[LOFAR constraints on FRB repetition \& environments]{LOFAR constraints on the repetition \& environments of CHIME FRBs}

\author[P. Chawla et al.]{Pragya Chawla$^{1,2}$\,\orcidlink{0000-0002-3426-7606},
Akshatha Gopinath$^{2}$\thanks{E-mail: a.gopinath@uva.nl}\,\orcidlink{0000-0002-1836-0771},
Ninisha Manaswini$^{2,3}$\,\orcidlink{0009-0005-1319-9586},
Cees Bassa$^{1}$\,\orcidlink{0000-0002-1429-9010},
Jason Hessels$^{2,1,4,5}$\,\orcidlink{0000-0003-2317-1446}, \newauthor
Vlad Kondratiev$^{1}$\,\orcidlink{0000-0001-8864-7471},
Daniele Michilli$^{6}$\,\orcidlink{0000-0002-2551-7554}, and
Ziggy Pleunis$^{2,1}$\,\orcidlink{0000-0002-4795-697X} 
\\
$^{1}$ASTRON, Netherlands Institute for Radio Astronomy, Oude Hoogeveensedijk 4, 7991 PD Dwingeloo, The Netherlands\\
$^{2}$Anton Pannekoek Institute for Astronomy, University of Amsterdam, Science Park 904, 1098 XH Amsterdam, The Netherlands\\
$^{3}$Max Planck Institute for Radio Astronomy, University of Bonn, Auf dem Hügel 69, D-53121, Bonn, Germany \\
$^{4}$Trottier Space Institute, McGill University, Montreal, Quebec, Canada \\
$^{5}$Department of Physics, McGill University, Montreal, Quebec, Canada \\
$^{6}$Laboratoire d’Astrophysique de Marseille, Aix-Marseille University, CNRS, CNES, Marseille, France
}

\date{Published online by Monthly Notices of the Royal Astronomical Society on  06 Nov 2025. DOI: 10.1093/mnras/staf1943}

\pubyear{2025}

\begin{document}
\label{firstpage}
\pagerange{\pageref{firstpage}--\pageref{lastpage}}
\maketitle

\begin{abstract}
The behaviour of fast radio bursts (FRBs) at radio frequencies $<400$\,MHz is poorly understood, with only two sources detected below 300\,MHz. We robustly characterise the 150-MHz activity of CHIME-detected FRB sources relative to their 600-MHz activity --- using their non-detection in 473 hours of archival observations from the LOFAR Tied-Array All-Sky Survey (LOTAAS), and in 252 hours of LOFAR observations of 14 repeating FRB sources -- the largest sub-300\,MHz targeted FRB campaign to date. We search the LOTAAS data for repeat bursts from 33 CHIME/FRB repeaters, 10 candidate repeaters and 430 apparent non-repeaters. Their non-detection yields a population-level statistical spectral index constraint of $\alpha_{s, 135\,\rm{MHz}/600\,\rm{MHz}}>-0.9$, indicating that FRB spectral indices are, on average, flatter than those of pulsars. From the targeted campaign, the prolific repeater FRB~20201124A shows $\alpha_s>0.55$, implying reduced low-frequency activity, unlike the typically negative $\alpha_\textrm{s}$ seen from FRBs at higher frequency bands. We explore free-free absorption in its circumburst environment as a cause of the non-detection at 150 MHz, and find that it is consistent with either a very young $\sim10$\,yr old supernova remnant; or a typical H\textsc{ii} region. Our simulations indicate that LOFAR2.0 can detect 0.3--9 FRBs per week, with up to 4 FRBs originating from redshifts $1<z<3$. Such detections will provide robust constraints on cosmological parameters due to their clean environments. Our results thus inform future low-frequency FRB searches through the limits we place on repetition rates, and show how even non-detections can place meaningful constraints on FRB circumburst environments. 
\end{abstract}




\section{Introduction}\label{sec:intro}
Fast radio bursts (FRBs) are energetic radio transients of microsecond to millisecond duration (see \citealt{cordes19} and \citealt{petroff19} for reviews). Over 3600 unique FRB sources have been discovered to date \citep{chime23}, with a small number securely localised to host galaxies (117; \citealt{nimmo25}). The localisations have confirmed the cosmological origin of most FRBs. Their physical origin is still largely unknown (see \citealt{platts19} for a summary of the proposed models). However, observations of FRB-like bursts from the Galactic magnetar SGR~1935+2154 suggest that at least some FRBs originate from magnetars \citep{bochenek20,chime20}.

About sixty FRB sources have been observed to repeat, i.e., about 3\% of the known population \citep{chime23}. The repeating and non-repeating sources exhibit significant differences in two observed burst properties, namely, temporal width and bandwidth \citep{scholz16,fonseca20,pleunis21}. The temporal widths of repeaters are, on average, larger and their bandwidths lower that those of the non-repeaters. Repeating FRB sources show complex morphology in their bursts. An often observed feature is downward-drifting (in frequency) band-limited sub-bursts within a burst, which is also known as the `sad-trombone' effect. More recently, instances of upward-drifting `happy-trombone' sub-bursts have been observed from repeating FRBs, although they are likely less common than their downward-drifting counterparts \citep{Zhou_2022,Faber_2024b, Wang_2025_arXiv, Zhang_2025_arXiv}.

These differences could be due to repeaters and non-repeaters having different progenitor sources, emission mechanisms and/or circumburst environments. Alternately, it is possible that all FRBs repeat and the one-off events are emitted by repeaters with extremely low repetition rates. This possibility is supported by population synthesis studies \citep{caleb19,james23}, analysis of volumetric occurrence rates \citep{ravi19b} and observational constraints on repetition rates \citep{chime23}. The observed differences in burst widths and bandwidths can be reconciled with the aforementioned hypothesis if the repetition rate is correlated with these properties \citep{connor20,chime23}. \citet{Hewitt_2022} found hints that FRB~20121102 shows both repeater- and non-repeater-like bursts. The energy distribution studies by \citet{Kirsten_2024} and \citet{Ould-Boukattine_2024} also suggest possible links between repeater- and non-repeater-like bursts, with the non-repeater-like bursts possibly being sampled from the high-energy tail of bursts from repeating sources.
 
The majority of FRBs have been discovered in the frequency range of 400--800\,MHz by the Canadian Hydrogen Intensity Mapping Experiment telescope (CHIME; \citealt{chime18}). FRB emission has been detected at frequencies as high as 8\,GHz in the case of one repeater (which would be as high as $\sim$10\,GHz in the source frame before redshifting, see, e.g., \citealt{Gajjar_2018}). However, there appears to a dearth of detectable FRBs at frequencies lower than 400\,MHz. Only three FRBs have been detected between 300--400\,MHz -- two as yet non-repeating, detected in all-sky surveys  \citep{parent20,bethapudi21}, and one repeating \citep{Bhusare_2025_ATel}. The inferred all-sky rate in this frequency range implies that FRBs exhibit either a flat spectrum, on average, or a turnover below 400\,MHz \citep{chawla17}. This conclusion is further supported by untargeted searches at lower frequencies. No FRBs have been detected in surveys conducted between 100--200\,MHz with telescopes such as the Low-Frequency Array (LOFAR;  \citealt{coenen14,karastergiou15}) and the Murchison Widefield Array (MWA; \citealt{tingay15,rowlinson16, Kemp2024}). We note that these surveys, with sky coverage ranging from 25 to $\sim$1000 sq. deg. and observing time of 10--1500\,h, had a much lower mapping efficiency than the ongoing higher-frequency search with CHIME. For comparison, the latter covers a sky area of $\sim$~250 sq. deg. and has been observing nearly continuously since operations began in 2018. We direct the reader to Table~1 in \citet{sokolowski24} for a recent comprehensive summary of non-targeted wide-field searches for low-frequency FRB emission. 

Several progenitor models propose that FRBs originate from young neutron stars, which can have a surrounding wind nebula and supernova remnant, and are in environments where star-formation has recently occurred (see, e.g., \citealt{connor16, margalit18}). A few repeating FRB sources have been confirmed to be embedded in extreme magneto-ionic environments with potential nebulae (persistent radio sources) surrounding them \citep{Marcote_2017, Niu_2022, Bruni_2025}. Absorption processes are typically at play in such environments and can introduce a spectral turnover \citep{rajwade17,ravi19}. Alternately, the turnover could be intrinsic to the emission mechanism. The synchrotron Galactic background, which scales as $\nu^{-2.55}$ \citep{haslam82}, can also significantly reduce the sensitivity of low-frequency observations at low Galactic latitudes. Additionally, scattering in the intervening ionised medium, expected to scale with frequency as $\sim\nu^{-4}$ (see, e.g., \citealt{lorimer04,Ocker_2022}), could render low-frequency FRB emission undetectable. As an example, FRB~20221219A has a scattering timescale of 19 ms at 1.4\,GHz \citep{faber24}, which would scale to 140 s at 150\,MHz. Such highly scattered bursts could not have been detected in the low-frequency searches conducted so far because the search timescales ranged up to a maximum of 1 s in the time domain \citep{coenen14} and 30 s in image plane searches \citep{rowlinson16}. Even if the searches probed such long timescales, scattering would smear the signal, effectively rendering it indistinguishable from radiometer noise.

With all these effects at play, targeted follow-up of repeating sources is arguably a more effective strategy to study the low-frequency emission of FRBs. Untargeted surveys need to observe large swaths of sky to have a reasonable chance of detection, making them computationally expensive. On the other hand, targeted observations can result in discovery of low-frequency emission using much less observational and compute time. Furthermore, the chance of detection in targeted observations can be maximised by observing known FRBs that have high repetition rates and low scattering times. 

This strategy has resulted in the LOFAR detections of FRB~20190212A (Gopinath et al., in prep.) at 150\,MHz, and the periodically active FRB~20180916B \citep{pastormarazuela21,pleunis21b,gopinath24} down to a frequency of 110\,MHz. More recently, FRB~20220912A \citep{Bhusare2022ATel} and FRB~20240114A \citep{Kumar_2024_ApJ, Panda2024} were detected down to frequencies of 300\,MHz by the uGMRT. Although emission from two repeaters has been detected between 110--190\,MHz, it is as yet unclear what fraction of FRB sources emit in this frequency range\footnote{There have been reports of FRBs detected at 111\,MHz using the Large Phased Antenna of the Lebedev Physical Institute with an observing bandwidth of 2.5\,MHz \citep{Fedorova_2019_ARep, Fedorova_2022_ATel,Tyul'bashev_2025_PASA}. Due to their small observing bandwidth, large untargeted search trials outside the localization region of the FRB sources, and/or low signal-to-noise, it is challenging to affirm the astrophysical origin of these detections.}. Targeted low-frequency searches for a large number of sources are required to understand the frequency dependence of FRB activity, particularly whether there exists a cutoff or turnover frequency for FRB emission. Detection (or a lack thereof) of such a turnover can constrain the emission mechanism of FRBs and characterise their local environments \citep{chawla20,pleunis21b}. Any sources detected at low frequencies can also be used for cosmological applications due to their clean environments (see, e.g., \citealt{macquart20}). Furthermore, targeted searches can allow for the determination of observing strategies for future surveys with telescopes such as LOFAR2.0\footnote{\url{https://www.lofar.eu/wp-content/uploads/2023/04/LOFAR2_0_White_Paper_v2023.1.pdf}} and SKA-Low \citep{dewdney09}.  

In this paper, we present results from LOFAR observations of CHIME-discovered FRB sources. We search for repeat bursts in archival LOFAR observations at the locations of CHIME repeaters and apparent non-repeaters and also conduct targeted observations of CHIME repeaters with LOFAR. Our paper is organised as follows. Section~\ref{sec:observations} presents the observational setup for the archival data from the LOTAAS survey and the targeted observations. Section~\ref{sec:analysis} describes the FRB search pipeline and its sensitivity to bursts from different FRB sources. Section~\ref{sec:constraints} describes the constraints on the frequency-dependent repetition of FRBs and their circumburst environments. Section~\ref{sec:future} presents rate predictions for future low-frequency FRB searches. We discuss our results and summarise our conclusions in Sections~\ref{sec:discussion} and \ref{sec:conclusions}.

\section{Observations}\label{sec:observations}
\subsection{LOFAR Tied-Array All-Sky Survey}\label{sec:lotaas}
\begin{figure}
\includegraphics[width=\columnwidth]{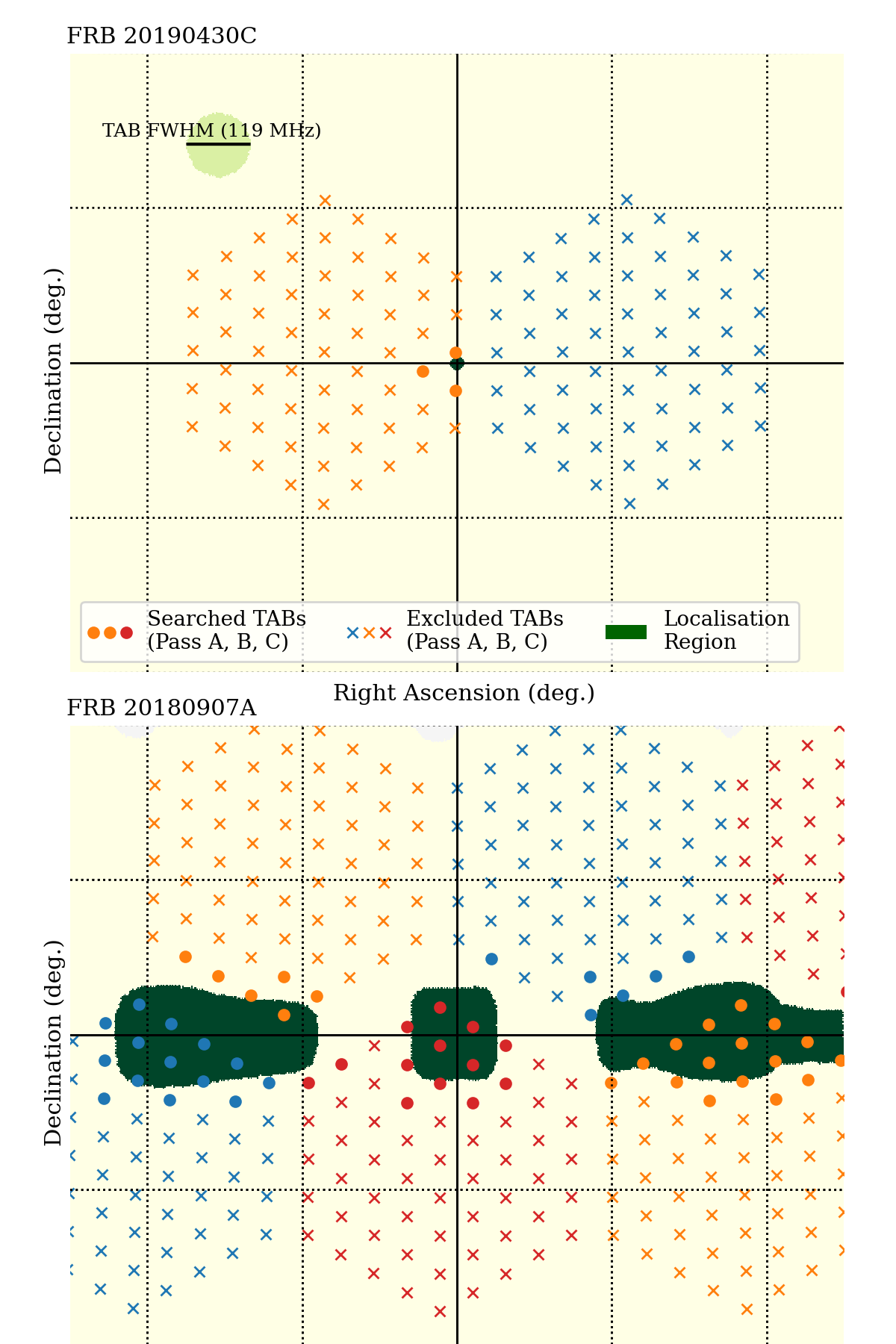}
    \caption{Examples of typical localization regions from CHIME baseband data (top) and real-time header information (bottom), and their overlap with the LOTAAS tied-array beams (TABs). We plot the LOTAAS TABs that overlap with the localisation regions of FRB~20190430C and FRB~20180907A and were searched for repeat bursts in this work. The localisation regions are plotted in green, with the positional uncertainty for FRB~20190430C increased by a factor of 5 for visual clarity. TABs which did not meet the overlap criterion (see \S\ref{sec:lotaas}) were excluded from the search and are marked with crosses. The marker colours indicate the survey pass that the TABs were observed in. The grid lines are spaced by 1$^\circ$.} 
    \label{fig:localisation}
\end{figure}

The LOFAR tied-array all-sky survey (LOTAAS; \citealt{sanidas19}) was conducted between 2012--2020. The survey was designed to search for radio pulsars and fast transients with the LOFAR telescope \citep{vanhaarlem13} and has discovered 74 pulsars to date. The survey was centered at 135.25\,MHz with a bandwidth of 31.64\,MHz using the dual-polarisation high-band antennas (HBAs) of the six central LOFAR stations (those located on the 400\,m-wide Superterp), the region of the LOFAR Core with the highest filling factor of stations. The entire northern sky ($\delta > 0^\circ$) was observed in three passes (termed A, B and C), with 651 pointings in each pass. Three station beams, referred to as sub-array pointings (SAPs), were formed for each observation. Signals from the HBAs were coherently summed to form 61 tied-array beams within each SAP \citep{stappers11,broekema18}. The tied-array beams (TABs) fill the central region of each SAP and their layout is shown in Figure~\ref{fig:localisation}. Twelve additional TABs were also formed within each SAP to enable pointing towards known pulsars that happen to lie within the SAP field-of-view. The survey setup is described in detail in \citet{sanidas19}.

For each TAB, total intensity was recorded for 2592 frequency channels in the range of 119--151\,MHz. The data were recorded with a time resolution of 491.52 $\mu$s for a duration of 1 h per pointing. The high time and frequency resolution result in a large data volume for each TAB (18 GB) even after re-quantisation from 32-bit floating points to 8-bit integers \citep{kondratiev16}. Therefore, in this work, we only process the data for the TABs at the location of CHIME/FRB sources, in order to make the computing requirements more manageable. Data for all TABs observed in the survey are available in the LOFAR Long Term Archive (LTA\footnote{\url{https://lta.lofar.eu/}}), with the exception of some pointings observed in the first survey pass. 
We investigate whether LOTAAS TABs overlap with the positions of 18 repeaters in the first CHIME/FRB catalog \citep{chime21} and 25 confirmed and 14 candidate repeaters presented by \citet{chime23}. For this analysis, we also consider two additional repeating sources, FRBs 20200120E and 20201124A, which are not included in the aforementioned catalogues but have been reported on separately due to their high repetition rates \citep{bhardwaj21, lanman22}.

Since it is as yet unclear whether non-repeating FRBs are truly one-off events, we also check for overlapping LOTAAS TABs for 453 apparent non-repeaters from the first CHIME/FRB catalog. Although there are 474 such sources in the catalogue, we exclude 12 sources which have been recently classified as confirmed or candidate repeaters by \citet{chime23}. An additional three sources were excluded due to unavailability of localisation information as they were detected in the far sidelobes of the CHIME primary beam. Furthermore, we exclude six sources, as their declinations were not within the range used for LOTAAS observations.

For the selected sample, we determine which LOTAAS TABs overlap with the localisation regions of the FRB sources. For this purpose, we only consider TABs in the central hexagonal grids, excluding the 12 TABs in each SAP that were pointed towards known pulsars. For the majority of the FRB sources, we use the localisation regions reported in the CHIME/FRB catalogues \citep{chime21,chime23}. However, three CHIME/FRB repeaters (FRBs 20180916B, 20200120E, and 20201124A) have been localised by the PRECISE (Pinpointing Repeating CHIME Sources with the EVN) program\footnote{\url{http://www.ira.inaf.it/precise/Home.html}} using the European Very Long Baseline Interferometry Network (EVN); for these we use the corresponding milliarcsecond localisations \citep{marcote20,kirsten22,nimmo22}. 

The FWHM of LOTAAS TABs varies with frequency, ranging from 0.32$^\circ$ at 151\,MHz to 0.41$^\circ$ at 119\,MHz. We assume the TAB extent to be the latter while determining the overlap with the localisation regions. A total of 33 repeaters and 430 non-repeaters had localisation regions that overlapped with LOTAAS TABs for which data were available in the LTA. The non-repeating sources are not well localised, often having three disjoint localisation regions (see bottom panel of Figure~\ref{fig:localisation}). An average of 40 TABs per source were required to effectively tile these localisation regions and were analysed using our FRB detection pipeline (see \S\ref{sec:pipeline}). Data for 80\% of all LOTAAS TABs were available in the LTA. Therefore, we analyse data corresponding to, on average, 80\% of the localisation regions of the CHIME FRB sources.

\subsection{Targeted Follow-up Observations}
Targeted observations of several CHIME/FRB repeaters were performed using LOFAR using the same strategy as used in \citet{pleunis21b} and \citet{gopinath24}. One TAB at the known source position, with a FWHM $\sim{3^{\prime}}$, was formed by coherently adding signals from the HBAs of up to 24 of the LOFAR Core stations. Dual-polarisation complex voltage data across 400 subbands, each 195.3125\,kHz wide, were recorded between 110--188\,MHz. The complex voltage data was used to make 3.05-kHz frequency channels across the entire 78.125-MHz wide band. Total intensity Stokes~I filterbank files were generated from summing the complex voltage data from the two orthogonal polarization hands. The data were channelized to 3.05\,kHz channels. Using \texttt{digifil} we averaged by a factor of 16 in frequency, after incoherently dedispersing the filterbank files at the known CHIME/FRB DM of the source to reduce dispersion smearing. With this we obtained filterbank files with a frequency resolution of 48.8\,kHz and a time resolution of 0.983 ms. Depending on the time between discovery and follow-up, the source positions and DMs used for the targeted observations were either: (1) the CHIME/FRB-reported values from an Astronomer's Telegram, (2) a CHIME/FRB VOEvent\footnote{\url{https://www.chime-frb.ca/voevents}}, (3) the CHIME/FRB baseband-localisation position \citep{Michilli_2023}, or (4) an internally communicated or publicly published milliarcsecond-precision position from the EVN-PRECISE program.

\begin{figure*}
\includegraphics[width=1.7\columnwidth]{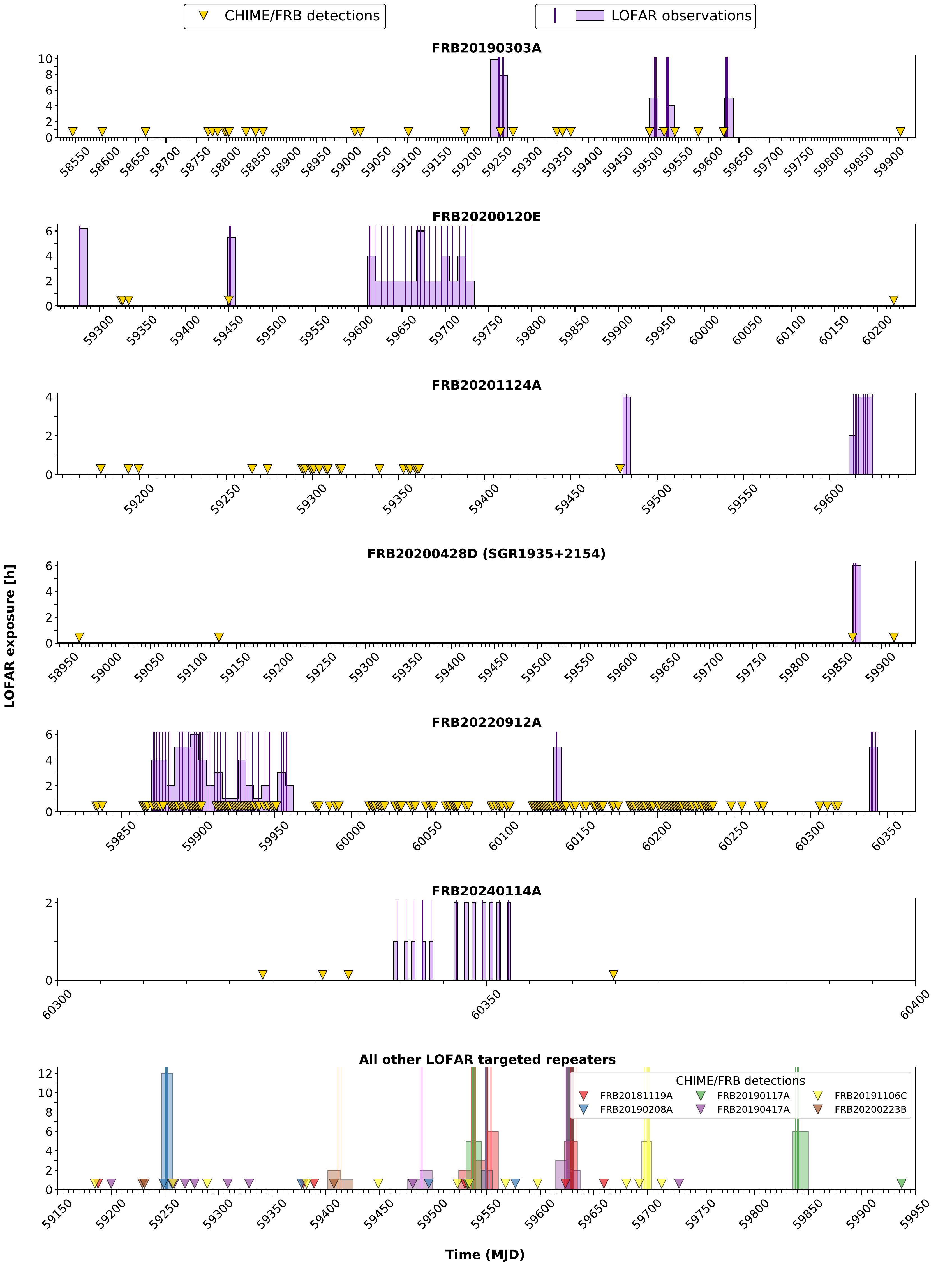}
    \caption{A timeline of targeted LOFAR exposures of CHIME/FRB repeaters, along with the CHIME/FRB burst detection times. Most individual LOFAR observations lasted 1--2 h on a single day. Each vertical line marker shows the time of a LOFAR observation session but the height of the line is arbitrary. The histogram bin heights show the exposure (in hours) but the bin widths are arbitrary. The figure emphasises how closely in time LOFAR observations were conducted relative to the most recent CHIME/FRB burst for each source. Detection times are marked up to one burst beyond the last LOFAR observation for each source. For the last panel which depicts observations of six repeaters, we prioritise summarising the LOFAR observations (histograms/vertical lines of different colours). Therefore, the first plotted CHIME burst detection (triangles of different colours) is an arbitrary amount of time before the first LOFAR observation of that source, and is not necessarily the first CHIME/FRB detection of the source. \newline The CHIME/FRB detection data are taken from their public database (\url{https://www.chime-frb.ca/}), which also includes bursts not accounted for in the rates used in this work (although this is not meant to be a complete burst sample). We use rates from the catalogue of CHIME/FRB repeaters \citep{chime23}, which only includes bursts detected up to May 1, 2021 (MJD~59335). For instance, the repeaters FRB~20220912A and FRB~20240114A were discovered after this cut-off date.}     \label{fig:timeline_targeted_repeaters}
\end{figure*}

The list of known repeating sources whose observations were conducted as a part of the LOFAR campaigns LT16\_008, LC17\_011, and LC20\_041 -- between July 2021 and February 2024 -- can be found in Table~\ref{tab:targeted_nondetections}. Note that only one source (FRB~20180916B) has a reported detection with LOFAR at 150\,MHz \citep{pleunis21, gopinath24}, and a second source (FRB~20190212A) that has been detected will be reported by Gopinath et al. (in prep.). Thus Table~\ref{tab:targeted_nondetections} exclusively lists sources that have not been detected so far in the above observation campaigns. The targeted observations were scheduled in response to either CHIME/FRB detections or other higher-frequency detections of the source. A visualization of this timeline is shown in Figure~\ref{fig:timeline_targeted_repeaters}.

\begin{table*}
\caption{Exposures per source from multiple campaigns of targeted LOFAR follow-up of CHIME/FRB repeaters between July 2021 and February 2024, for which no bursts were detected. Fluence thresholds ($F_{\textrm{th,L}}$) for the LOFAR observations are calculated using Equation~\ref{eq:threshold}, based on burst widths derived using CHIME/FRB measurements at 600\,MHz. We quote the 90\% confidence upper limits on the LOFAR burst rates ($\lambda_\textrm{L}$) at 150\,MHz and the 90\% confidence lower limits on the statistical spectral index $\alpha_{s\textrm{, 150\,MHz/600\,MHz}}$.}
\label{tab:targeted_nondetections}
\begin{threeparttable}[b]
    \centering
    \begin{tabular}{ccccc}
    \hline
    Source        & Exposure (h)  & $F_{\textrm{th,L}}$ (Jy ms)\tnote{a} & $\lambda_\textrm{L}$ (h$^{-1}$) & $\alpha_{s, 150\,\rm{MHz}/600\,\rm{MHz}}$\\
    \hline
    FRB~20181119A     & $16.0$            & 142         & ${<0.14}$    &   $>-1.70$  \\
    FRB~20190117A     & $11.0$            & 218         & ${<0.21}$    &   $>-1.63$  \\
    FRB~20190208A     & $12.7$            & 159         & ${<0.18}$    &   $>-1.93$  \\
    FRB~20190303A     & $32.7$            & 143         & ${<0.07}$    &   $>-1.39$  \\
    FRB~20190417A     & $9.0$             & 120         & ${<0.26}$    &   $>-1.67$  \\
    FRB~20191106C     & $5.0$             & 482         & ${<0.46}$    &   $>-2.88$  \\
    FRB~20200120E     & $47.7$            & 21          & ${<0.05}$    &   $>-0.95$  \\
    FRB~20200223B     & $3.0$             & 157         & ${<0.77}$    &   $>-1.95$  \\
    FRB~20200428D     & $6.0$             & 287         & ${<0.38}$    &   $>-1.44$  \\
    FRB~20201124A     & $14.0$            & 11          & ${<0.16}$    &   $>+0.55$  \\
    FRB~20220912A     & $61.0$    & 253         & ${<0.04}$    &    -        \\
    FRB~20240114A\tnote{b}    & $19.0$    & 76          & ${<0.12}$    &   $>-1.10$   \\
    FRB~20210822E\tnote{c}    & $10.0$    &      -      &      -       &      -      \\
    FRB~20220316A\tnote{c}    & $5.0 $    &      -      &      -       &      -      \\
    \hline
    \end{tabular}  
    \begin{tablenotes}
        \item [a] {Fluence thresholds are calculated using expected burst widths at 150\,MHz, calculated using intrinsic widths and scattering timescales measured by \citet{chime23}. For FRB~20220912A and FRB~20240114A, we used intrinsic width and scattering measurements from uGMRT observations at 650\,MHz \citep{Bhusare2022ATel} and from CHIME/FRB observations at 600\,MHz \citep{Shin_2025_arXiv}, respectively.}
        \item [b] {FRB~20220912A, a prolific repeater, was followed up by LOFAR at 150\,MHz upon reports of detections from various telescopes including CHIME/FRB \citep{McKinven_2022_ATel}. However, it has no reported CHIME/FRB burst counts or exposure information. As a result, we are unable to perform the analysis to calculate a statistical spectral index for this source.}
        \item [c] {FRB~20210822E and FRB~20220316A are as yet unconfirmed repeating sources. We targeted them based on the CHIME VOEvents Service, using the sky position and DM information to determine the association of three or more events with each of these two sources. Hence, we only report the exposure hours.}
    \end{tablenotes}
\end{threeparttable}
\end{table*}

\section{Analysis}\label{sec:analysis}
\subsection{FRB Search Pipeline}\label{sec:pipeline}

\textit{LOTAAS:}\newline
The analysis pipeline, based on the \texttt{PRESTO} software package \citep{ransom01}, was run on the Dutch National Supercomputer Snellius. Data for the selected LOTAAS TABs were downloaded from the LOFAR LTA. The initial stages of our pipeline are the same as those used by \citet{sanidas19} for searching the LOTAAS data for pulsars. In short, we first mask time samples and frequency channels containing radio-frequency interference (RFI) using the \texttt{rfifind} tool in \texttt{PRESTO}. The data were then dedispersed to a large number of trial DMs up to a maximum of 3020 pc~cm$^{-3}$. Although the DMs of the CHIME sources are known, we search over a range of trial DMs to maximise our chances of serendipitously discovering new FRB sources in these beams. Additionally, many FRBs are known to have multiple band-limited components that drift downwards in frequency \citep{hessels19}, which can look like uncorrected residual dispersion, especially if combined with temporal widening of the bursts by scattering. In such cases, the signal-to-noise maximizing DM can be greater than the true DM of the bursts (see, e.g., burst B25 in \citealt{gopinath24}). 

The dedispersed time series were searched for single pulses using \texttt{PRESTO}’s matched filtering routine, \texttt{single\_pulse\_search.py}. A fraction of the TABs in our sample were searched for single pulses with widths $<$ 100 ms up to a DM of 550 pc~cm$^{-3}$ in the initial survey processing performed by \citet{sanidas19}. We improve on the initial processing by extending the trial DM range to 3020 pc~cm$^{-3}$ (over $\sim$14000 trial DMs) and searching for pulse widths up to 500 ms. The choice of the maximum trial DM is based on the highest-DM event in the CHIME/FRB catalog (3015 pc~cm$^{-3}$), while the rationale for the maximum width is discussed in \S\ref{sec:sensitivity}. 

The output of the single-pulse search is processed by a grouping algorithm from the software package \texttt{L-SPS} \citep{michilli18}. Due to the large number of closely spaced trial DMs, any pulse (astrophysical or RFI) is detected in multiple dedispersed time series. After \texttt{L-SPS} groups the single-pulse events that are close in time and DM, we employ a machine-learning classifier, \texttt{FETCH} \citep{agarwal20}. We use \texttt{FETCH} to effectively identify faint (S/N $>$ 7) astrophysical signals among the large number of false positives generated by RFI. FETCH assigns each candidate a probability of being astrophysical based on its dynamic spectrum and the DM-time image (commonly referred to as the ``bow-tie'' plot). We visually inspect the dynamic spectra of all candidates that are assigned a probability above 50\% by any of the \texttt{FETCH} deep-learning models \texttt{a} to \texttt{h}. 

About 20\% of the LOTAAS TABs had a large number of candidates ($\sim$10$^3$), making it prohibitive to generate dynamic spectra for all of them. For these beams, we employ two additional methods to reduce the number of candidates requiring visual inspection. First, we only inspect candidates with DMs in a narrow range of 5 pc~cm$^{-3}$ around the known FRB DM. Second, we vary the S/N threshold for visual inspection based on the candidate width. Therefore, for these beams, the S/N threshold of 7 for inspection is only applicable for candidates with widths $< 100$ ms. Since the majority of the false positives had larger widths, candidates with widths $> 400$ ms were only inspected if their S/N was higher than 12. A S/N threshold between 7 and 12 was used for candidates with widths of 100--400 ms.
\newline
\newline
\textit{Targeted observations:}\newline
For the targeted observations, a search strategy similar to the one employed by \citet{gopinath24} was used. We mask the time and frequency bins containing RFI using \texttt{rfifind}, and further replace these bins with Gaussian noise corresponding to local statistics around these bins. The searches were, in general, performed on filterbank files with a time resolution of $0.983$ ms -- unless there were known constraints on the scattering timescale from detections at higher frequencies. In cases where the scattering timescale limits -- when extrapolated to 150\,MHz ($\tau_{s,150\,\textrm{MHz}}$) using an assumed frequency ($\nu$) dependent scaling of $\tau_s \propto \nu^{-4}$ -- were less than a few milliseconds, additional searches were conducted at higher time resolutions (up to $81\,\mu$s). The higher-time-resolution searches were enabled by coherently dedispersing the data using \texttt{cdmt} \citep{bassa2017a}, with the resulting output filterbank file still having the same time and frequency resolution as the lower-resolution search (with incoherent dedispersion), but with reduced dispersion smearing. This applied to the search for bursts in the observations of FRBs 20200120E, 20201124A, 20200428D, and 20220912A. However, these higher-time-resolution searches did not yield any positive burst candidates for any of the sources. Therefore, for the sake of simplicity and consistency, we only consider the searches at $0.983$ ms resolution for the rest of the analysis. 

All the targeted observations were additionally searched by creating filterbank files of 4 overlapping halves and 8 overlapping quarters of the 78.8-MHz LOFAR HBA band to increase sensitivity towards narrow-band bursts. These fractional-band filterbanks were then searched using the same pipeline as described above. Later, these subbanded searches were extended to 15 overlapping 12.5\,MHz-wide searches, with approximately 12-15\% of all targeted observations being searched this way. Since there were no detections of astrophysical events above a S/N of 7, we do not consider these sub-banded searches for further analysis, while still noting that they helped increase the sensitivity of our searches.

After RFI flagging, \texttt{single\_pulse\_search.py} was used to search for bursts of widths up to $150$ ms, using 400 DM trials around the known CHIME/FRB DM, in steps of 0.1\,pc~cm$^{-3}$. Further, the output of the single-pulse search was filtered for potential bursts using \texttt{L-SPS} and FETCH as described above, before a manual visual inspection of all the candidates above a S/N threshold of 7 and a FETCH median-probability higher than 50\% among its deep learning models \texttt{a} to \texttt{h}\footnote{In the LOFAR searches conducted for bursts from FRB 20180916B \citep{gopinath24, pleunis21}, we noticed that some FETCH models had a higher false positive rate than others, while a different subset of models had a higher false negative rate. In order to mitigate both these sources of error, we adopted the median probability across all models as the preferred metric for selecting candidates for visual inspection in the targeted observations.}.The DM-time bow-tie plots that the FETCH pipeline uses for classification were modified before the classification step to zoom in along the DM axis to DMs ranging between $[0.9\times \textrm{DM}_{\textrm{known}}, 1.1\times {\textrm{DM}_{\textrm{known}}}]$, instead of the default $[0, 2 \times{\textrm{DM}_{\textrm{known}}}]$, where $\textrm{DM}_{\textrm{known}}$ corresponded to the CHIME-reported DM for the source.

\subsection{Search Sensitivity}\label{sec:sensitivity}
We determine the minimum detectable fluence for the LOFAR searches using the radiometer equation (see, e.g., Equations 15 and 16 in \citealt{cordes03} from which the below can be derived), 
\begin{equation}\label{eq:threshold}
F = \frac{\textrm{S/N} \ (T_\textrm{rec} + T_\textrm{sky})}{G}\sqrt{\frac{W_\textrm{b}}{n_\textrm{p} \Delta\nu}}.
\end{equation}
\newline
\newline
\textit{LOTAAS:} \newline
For LOTAAS observations, the receiver temperature ($T_\textrm{rec}$) and telescope gain ($G$) are set to 360 K and 1.7 K/Jy, respectively, following \citet{sanidas19}. The number of summed polarisations ($n_\textrm{p}$) is 2 and the observing bandwidth ($\Delta\nu$) is 31.64\,MHz. We estimate $T_\textrm{sky}$ at the location of each CHIME FRB using the 408\,MHz all-sky map of \citet{remazeilles15}. The temperature is then scaled to the centre frequency of LOTAAS (135\,MHz) using a spectral index of $-2.55$ for Galactic synchrotron emission \citep{haslam82}. \citet{Price_2021_RNAAS} find that global sky models (GSM) provide more accurate $T_\textrm{sky}$ estimates as they account for frequency- and position-dependent variations in the spectral index. In the 110--188 MHz band, the \citet{haslam82} map differs from the GSM by $< 15$\% for most CHIME/FRB sources, with variations averaging $\sim$2\% for sources off the Galactic plane \citep{Price_2021_RNAAS}. Since these differences are below the typical $\sim20$\% uncertainties in fluence thresholds (dominated by parameters other than $T_\textrm{sky}$), adopting GSM-based estimates will not significantly impact our results. 

The search sensitivity also depends on the minimum detectable signal-to-noise ratio, S/N$_\textrm{th}$, for the analysis pipeline. As discussed in \S\ref{sec:pipeline}, we used a S/N threshold of 7 for about 80\% of the LOTAAS TABs, with the remaining beams having a threshold dependent on the expected broadened pulse width ($W_\textrm{b}$) of the FRB source being searched. We set S/N$_\textrm{th}$ to be the weighted average of the two S/N thresholds for the purpose of computing a single value of $F_\textrm{th,L}$ across all TABs searched for each CHIME FRB. The weights are equal to the fraction of the data analysed with each threshold. 

The broadened pulse width, $W_\textrm{b}$, is the quadrature sum of the sampling time, the dispersive smearing within each frequency channel as well as the intrinsic width and scattering timescale of the CHIME sources. Here we set the sampling time to be 0.983 ms as the data are downsampled by a factor of 2 prior to running the search pipeline. The intra-channel dispersive smearing at 135\,MHz is greater than 0.983 ms for DMs $>25$ pc~cm$^{-3}$. The downsampling thus reduces the processing time without compromising the search sensitivity for the FRBs in our sample, all of which have DMs $>100$ pc cm$^{-3}$. 

To calculate $W_\textrm{b}$, we use the intrinsic width and the scattering timescale ($t_\textrm{scatt}$) for each FRB source from the corresponding CHIME/FRB catalogues. In case of repeaters, we set the intrinsic width to be equal to the mean value for all repeat bursts observed with CHIME. Assuming that the scattering structures along the line-of-sight to these sources do not change significantly in the time between detections, we adopt the strongest constraint (i.e, the shortest measured scattering time) among all repeat bursts as the scattering time for each source. If the reported width and scattering time measurements are upper limits, we assume the 95\% confidence upper limit to be the measured value. The scattering timescales are reported for a frequency of 600\,MHz, which we scale to 135\,MHz by assuming a power-law index of $-4$ for the frequency dependence. This scaling from 600\,MHz makes the scattering times at 135\,MHz larger by a fixed factor of 390. 

For some repeating FRBs, follow-up observations with other telescopes have provided stronger constraints on their scattering timescale. For FRB~20180916B, we use the 95\% confidence upper limit reported by \citet{chawla20} of 1.7 ms at 350\,MHz. For FRB~20201124A, \citet{lanman22} measure a scattering time of 8 ms at 600\,MHz. However, they mention the possibility that unresolved downward-drifting sub-bursts are being misinterpreted as scattering by their fitting algorithm. This possibility is supported by the scintillation bandwidth measured for this source by \citet{main22}, which implies a much lower scattering timescale of 2\,$\mu$s at 600\,MHz. Therefore, we adopt a value of 2\,$\mu$s for the scattering time for FRB~20201124A when calculating the LOFAR sensitivity to the source.   

For the FRBs in our sample, the expected scattering timescales at 135\,MHz range from 0.9 ms to 35 s. However, \texttt{single\_pulse\_search.py} (see \S\ref{sec:pipeline}) searches for bursts through convolution of each dedispersed time series with boxcar functions of widths up to 500 ms. Therefore, if a CHIME/FRB source with $t_\textrm{scatt, 135\,MHz} > 500 \textrm{ms}$ emits a repeat burst during LOFAR observations, the S/N of the detected burst will only be derived from a fraction of the burst fluence, i.e., the area of the brightest 500-ms time chunk within the burst. To account for the reduction in sensitivity due to this effect, we need to divide the fluence threshold derived in Equation~\ref{eq:threshold} by this fraction (hereafter referred to as $f_\textrm{boxcar}$).

\begin{figure}
\includegraphics[width=\columnwidth]{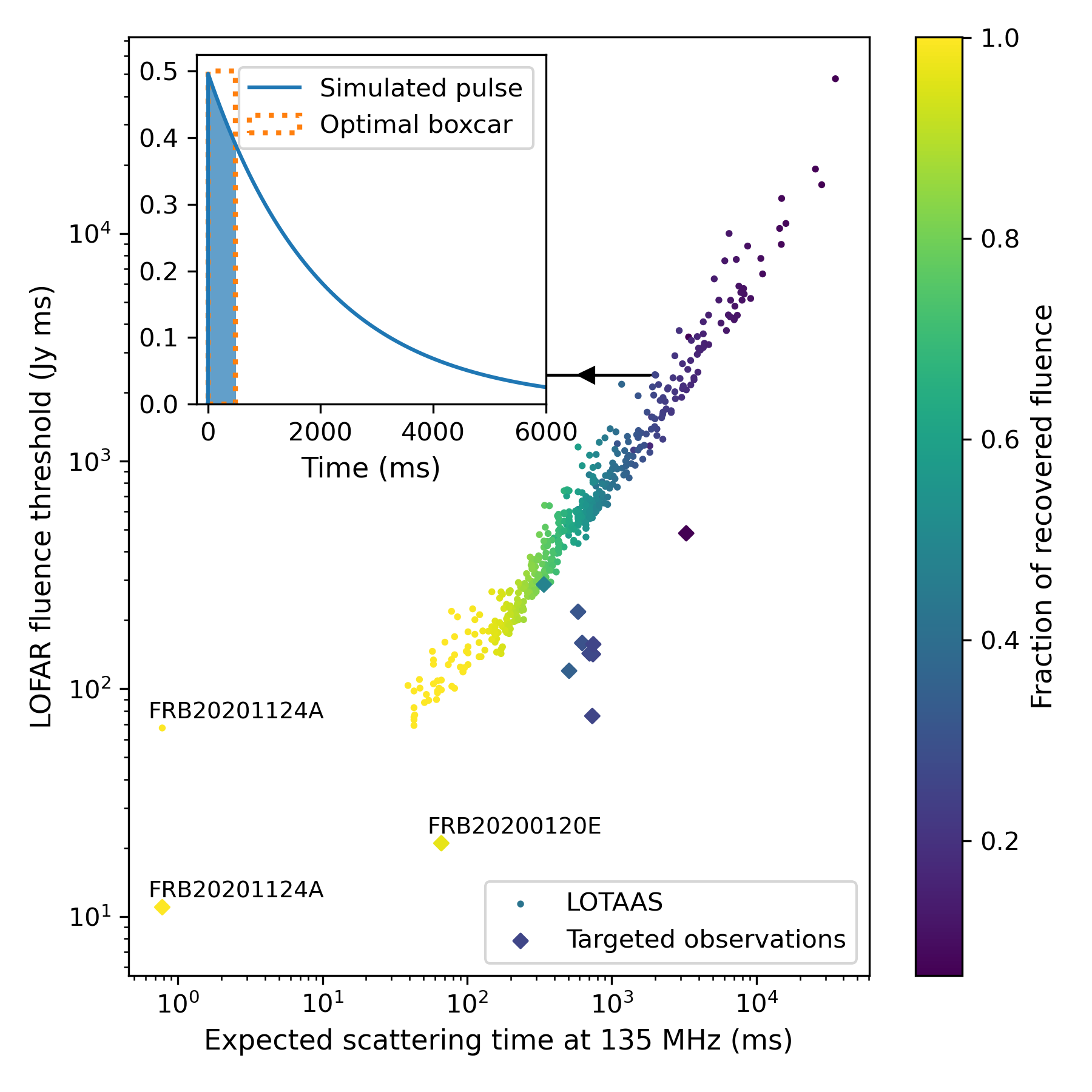}
    \caption{Fluence threshold of LOFAR observations for the CHIME FRB sources plotted as a function of the expected scattering time at 135\,MHz. The thresholds account for the fraction of fluence of scattered bursts that can be recovered with our search pipeline. This fraction, $f_\textrm{boxcar}$ (indicated by the marker colour), is derived by simulating a burst with a width and scattering time the same as that measured for the corresponding source. The inset illustrates that for a burst with a scattering timescale of 2 s, only 20\% of the fluence (indicated by the shaded blue region) can be recovered with the widest boxcar in our search pipeline.}
    \label{fig:fluenceth}
\end{figure}

For each source in our sample, we determine $f_\textrm{boxcar}$ by first simulating a pulse as the convolution of a Gaussian function with a one-sided exponential function \citep{mckinnon14}. In doing so, we assume the CHIME-measured intrinsic width to be the standard deviation of the Gaussian profile and $t_\textrm{scatt}$ as the decay timescale of the exponential. The simulated pulse is convolved with boxcar functions of widths ranging from 2 to 500 ms, the same as those used by our pipeline. We obtain the fraction of the burst fluence recovered by each boxcar by dividing the maximum value of the convolution by the total area under the pulse. The highest value of this fraction, i.e. that for the optimal boxcar, is adopted as the correction factor $f_\textrm{boxcar}$. This factor could, in principle, be obtained analytically as a function of the ratio between the scattering time and the intrinsic pulse width. However, we choose to obtain it empirically, as described, so as to retain the functional form of \citet{mckinnon14} employed by the CHIME/FRB collaboration in their burst-fitting process \citep{fonseca20}. The derived correction factor and fluence threshold for each source, as a function of its scattering timescale, is shown in Figure~\ref{fig:fluenceth}. To avoid accounting for the reduction in sensitivity due to scattering twice, the value of $W_\textrm{b}$ in Equation~\ref{eq:threshold} is capped at 500 ms. While this correction increases the robustness of the sensitivity calculation, it does not significantly affect the constraints presented in \S\ref{sec:alpha_s} and \S\ref{sec:env} as the dominant contributor to those constraints are sources with low scattering times, for which the fluence thresholds are unaffected by this correction. 
\newline
\newline
\textit{Targeted observations:} \newline
The searches in the targeted follow-up observations had a S/N threshold of 7, a sampling time of $0.983$ ms, and a \texttt{single\_pulse\_search.py} maximum boxcar width of $150$ ms across all observations. The effective pulse width $W_{b}$ and the boxcar-determined fractional width $f_{\textrm{boxcar}}$ are calculated in the same way as for the LOTAAS observations as described above, except that the maximum boxcar width is now $150$ ms instead of $500$ ms. The number of LOFAR HBA cores for any given (typically $\sim{1}$-h long) source observation tended to vary between 20 and 24 tiles, which alters the effective area of the telescope, and is taken into account in our calculation. The LOFAR response is strongly dependent on elevation since LOFAR is a phased-array antenna; this is taken into account for $T_\textrm{rec}$ and $G$ values, as detailed in \citet{kondratiev16}. The gain calculation uses the Hamaker beam model \citep{Hamaker2011}, implemented using the \texttt{mscorpol}\footnote{\url{https://github.com/2baOrNot2ba/mscorpol}} package. The receiver temperature $T_\textrm{rec}$ varies both with antenna pointing and with frequency. There are grating lobes that appear at the higher frequencies of the band when pointing away from the zenith with a decreased effective area, and the effective impedance of the antenna elements varies at the lower frequencies of the band. This is encapsulated by a polynomial fit to measurements from \citet{Wijnholds2011}. $T_{\textrm{sky}}$ at the FRB positions from the 408-MHz all-sky map of \citet{haslam82} is scaled to a central frequency of $150$\,MHz using a spectral index of $-2.55$, and dominates $T_{\textrm{sys}}$. The bandwidth for the targeted observations was 78.8\,MHz centered around 150\,MHz. Most of the RFI observed in this band is in the top $\sim15$\,MHz of the band, typically totalling to $\sim5$\% of the entire bandwidth. We do not find it necessary to account for this small fraction of RFI occupation in reporting our observing bandwidth, and hence the sensitivity limit. In relation to the LOTAAS survey searches with a bandwidth of 32\,MHz, this is a higher observing bandwidth. Considered along with the higher central frequency than the LOTAAS observing band, the effect of the steep scaling of scattering and sky temperature at such low frequencies is relatively mitigated. Both these factors increase the sensitivity of the targeted observations. All other parameters for the calculation of the fluence threshold are as previously described. Given that we performed sub-banded searches for all the targeted sources \citep[since most repeating FRB bursts are band-limited]{hessels19}, and used coherent dedispersion to perform higher-time-resolution searches for many of the prolific repeaters, our quoted sensitivity limits are very conservative (up to a factor of $\sim3$). The fluence threshold, $F_{\textrm{th}}$, is then calculated as in Equation~\ref{eq:threshold}, and the value for each source is reported in Table~\ref{tab:targeted_nondetections}. We report the fluence thresholds averaged across all observations of a given source, with the value for each observation using the band-averaged gain and $T_\textrm{sys}$. 

\subsubsection{Performance of FETCH}
In addition to the fluence, the detection of an FRB by our pipeline depends on whether FETCH assigns it a probability above 50\% of being astrophysical. We evaluate the performance of FETCH through an analysis of LOTAAS observations of eight Galactic radio pulsars. We select these pulsars from the ATNF Pulsar Catalogue\footnote{\url{https://www.atnf.csiro.au/research/pulsar/psrcat/}} \citep{manchester05}, choosing sources located within 2$^\circ$ of the centres of LOTAAS SAPs. The DMs of the selected pulsars were in the range 21--45\,pc~cm$^{-3}$ and their pulses spanned widths (measured at 50\% of the pulse peak) in the range 13--73\,ms. The data were then analysed using the same pipeline as for the FRB search. A total of 350 pulsar pulses were detected in these observations, with 200, 90 and 60 pulses having S/Ns in the ranges between 7--10, 10--20 and $>20$, respectively. 

We find that FETCH identifies all these pulses as being astrophysical,with at least one of the FETCH models assigning them probabilities $>50$\%. However, not all models assign probabilities $>50$\% to every pulsar pulse in these data, with large variations between models for pulses with 7 $<$ S/N $<$ 10. Therefore, for the FRB searches in this work, we do not rely on a single FETCH model. As described in \S \ref{sec:pipeline}, we either use the median probability assigned by all FETCH models as the selection criterion (for the targeted observations) or inspect candidates assigned a probability $>50$\% by any FETCH model (for the LOTAAS observations).

\section{Results}\label{sec:constraints}
We visually inspected the dynamic spectrum of all FRB candidates classified by FETCH as being astrophysical. The majority of flagged candidates had characteristics consistent with RFI, such as occupying only a few channels in frequency, or associated with baseline variations on longer timescales. We observed some faint candidates (S/N $<7.5$) that showed no obvious characteristics of RFI. However, their DMs varied by more than 15\,pc~cm$^{-3}$ from those measured for the FRB sources which they are co-located with. Therefore, we conclude that no repeat burst with a fluence greater than the estimated threshold for the corresponding FRB source (see Figure~\ref{fig:fluenceth}) was present in the LOTAAS pointings. It could be that the aforementioned faint candidates are truly astrophysical. However, in the absence of a confirmed high-S/N detection of an FRB source at the same DM and sky location, we cannot confirm the astrophysical origin of these candidates. Given that these bursts are only marginally above the detection threshold, the constraints on FRB repetition rates, spectra and circumburst environments, that we present below, would remain valid even if the bursts are later found to be astrophysical.

Based on the non-detection in the LOTAAS searches, we determine upper limits on the repetition rates of the CHIME/FRB sources at 135\,MHz. Using the duration of each LOTAAS observation (1 h), we derive a 90\% confidence upper limit on the rate for each source, $\lambda_\textrm{L} < 2.3$ bursts h$^{-1}$. This rate corresponds to the fluence thresholds plotted in Figure~\ref{fig:fluenceth}. 
The upper limit has been derived assuming that FRB repetition follows Poissonian statistics. Moreover, the upper limits are only valid for 24 repeaters and 114 non-repeating sources for which the available LOTAAS TABs covered the entire localisation region. This is because the LOTAAS exposure to a source could be zero if the analysed beams did not cover the entire localisation region.

We detect no bursts in the targeted observations reported here above a S/N threshold of 7. The 90\% confidence upper-limits on the rates of repeating sources observed through the targeted observations are tabulated in Table~\ref{tab:targeted_nondetections}, along with the corresponding fluence threshold limits. The most stringent burst rate limits are for FRB~20220912A ($\lambda_L < 0.04 \ \textrm{h}^{-1}$), FRB~20200120E ($\lambda_L < 0.05 \ \textrm{h}^{-1}$), and FRB~20190303A ($\lambda_L < 0.07 \ \textrm{h}^{-1}$), all of which had among the highest targeted observing hours spent on them.

The non-detection of repeat bursts in these observations could be due to several reasons. As for the non-repeating FRB sources, it is possible that these FRBs are truly one-off events. Considering that two repeaters exhibit a periodic modulation in their activity \citep{chime20b,rajwade20}, it could be that some of the repeating sources in our sample were not active at the time of LOFAR observations. Although repeaters are also highly variable in their burst rate, it is unlikely that all sources in our sample are inactive at the time of our observations, we consider other explanations for the non-detections. First, the burst rate and/or brightness could be intrinsically lower at low frequencies due to the emission mechanism. Alternately, propagation effects such as scattering in the intervening medium and absorption in the circumburst environment could render low-frequency bursts undetectable.  

In light of the aforementioned scenarios, the non-detection of bursts in our observations allows us to constrain the properties of the circumburst environments of FRBs and the frequency dependence of their emission. In \S\ref{sec:alpha_s}, we constrain the statistical spectral index ($\alpha_\textrm{s}$), a property which characterises the observed reduction in FRB rate at low frequencies, and is agnostic to the phenomenon that causes the reduction. In \S\ref{sec:env}, we investigate absorption in the circumburst environment as a potential cause of the non-detection of low-frequency emission from highly active repeating FRBs. We then constrain the electron density and temperature in the circumburst environments of these sources.

\subsection{Constraints on FRB Spectra}\label{sec:alpha_s}
We characterise the frequency dependence of FRB activity using the statistical spectral index ($\alpha_\textrm{s}$), following \citet{houben19}. The index characterises the assumed power law relating the normalisation ($A$) at different frequencies of the differential energy distribution of an FRB source, 
\begin{equation}\label{eq:energydist}
\frac{dN(\nu)}{dE_\nu} = A(\nu) E^{\gamma_\textrm{src}}_{\nu}.
\end{equation}
Here, we assume that the power-law index, $\gamma_\textrm{src}$, characterising the differential energy distribution is different for each FRB source and ${dN(\nu)}/{dE_\nu}$ is the number of bursts with spectral energies in the range [$E_\nu$, $E_\nu + dE_\nu$] \,erg Hz$^{-1}$.

The above parameterisation dictates that the repetition rates for LOFAR and CHIME, $\lambda_\textrm{L}$ and $\lambda_\textrm{C}$, defined at the respective central observing frequencies ($\nu_\textrm{L}$ and $\nu_\textrm{C}$) are related in the following manner, 
\begin{equation}\label{eq:alpha_s}
\frac{\lambda_\textrm{L}}{\lambda_\textrm{C}} = \bigg(\frac{\nu_\textrm{L}}{\nu_\textrm{C}}\bigg)^{-\alpha_\textrm{s} \gamma_\textrm{src}} \bigg(\frac{F_{\textrm{th,L}}}{F_{\textrm{th,C}}}\bigg)^{\gamma_\textrm{src}+1}.
\end{equation}

In this work, we use $\alpha_\textrm{s}$ to characterise the frequency dependence of FRB repetition as opposed to an instantaneous spectral index characterising the variation in burst intensity with frequency. This is because the instantaneous spectral index can vary greatly between repeat bursts from the same source \citep{scholz16}. Moreover, repeating FRBs are known to emit bursts with bandwidths as low as 20\,MHz \citep{hessels19,gopinath24}, for which an instantaneous spectral index is not physically meaningful. Therefore, we instead measure the statistical spectral index which characterises the frequency dependence of the rate at which narrow-band bursts of a particular energy are emitted. 

To constrain the statistical spectral index for each repeating FRB source, we solve Equation~\ref{eq:alpha_s} using the repetition rate ($\lambda_\textrm{C}$) and 90\% confidence completeness threshold ($F_{\textrm{th,C}}$) measured by \citet{chime23}. We use the values measured in this work for the corresponding quantities at LOFAR frequencies ($\lambda_\textrm{L}$ and $F_{\textrm{th,L}}$). Therefore, we can only constrain $\alpha_\textrm{s}$ for 22 repeaters in the LOTAAS survey, and for 11 repeaters in the targeted observations, which have their repetition rates constrained using both CHIME and LOFAR. Some of these repeaters are present in both the LOTAAS and targeted observation samples. 

\begin{figure}
\includegraphics[width=\columnwidth]{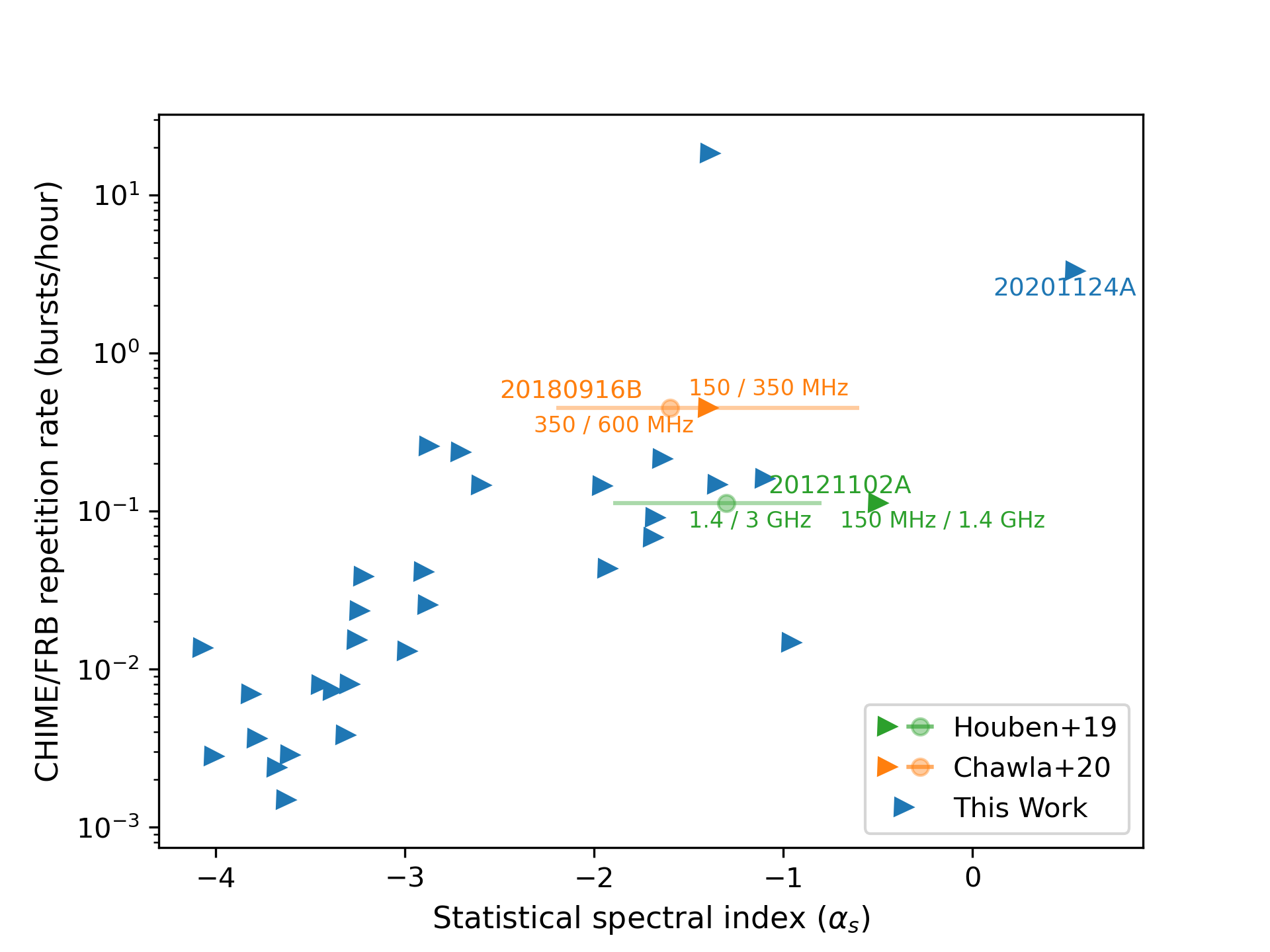}
    \caption{Constraints on the statistical spectral index ($\alpha_\textrm{s}$) of repeating FRBs derived by comparing burst rates in different frequency ranges. The lower limits are derived by comparing rate upper limits in LOFAR observations (110--190\,MHz) with the rate measured at higher frequencies. The frequency ranges corresponding to measurements by \citet{houben19} and \citet{chawla20} are indicated on the plot. The CHIME/FRB repetition rates are all scaled to a fluence threshold of 5 Jy~ms, assuming the power-law index of the differential energy distribution, $\gamma_\textrm{src} = -2.5$. For sources with available constraints from both LOTAAS and targeted observations, the stronger of the two constraints is plotted here.}
    \label{fig:indexpersrc}
\end{figure}

Since the measured repetition rates have significant uncertainties, we sample 1,000 values each for $\lambda_\textrm{L}$ and $\lambda_\textrm{C}$ within their reported 90\% confidence intervals. We sample $\gamma_\textrm{src}$ from a Gaussian distribution with a mean of $-2.5$ and a standard deviation of $0.5$. These parameter values are chosen to ensure the simulated power-law indices span the range measured for repeating FRBs ($-1.6$ to $-4.6$; \citealt{li21,lanman22}). For each of the 1,000 simulated sets of $\lambda_\textrm{L}$, $\lambda_\textrm{C}$ and $\gamma_\textrm{src}$, we solve for $\alpha_\textrm{s}$ and report the 10th percentile of the resulting distribution as the 90\% confidence lower limit in Figure~\ref{fig:indexpersrc}. The $\alpha_\textrm{s}$ for the targeted FRB sources are also tabulated in Table~\ref{tab:targeted_nondetections}.

The strongest constraint from the LOTAAS observations, $\alpha_\textrm{s} > -0.9$, is derived from the highly active source, FRB~20201124A. However, if the source only began bursting around the time of its discovery, this constraint may not be valid as the archival LOTAAS observations were conducted prior to that epoch. In contrast, all constraints derived from targeted follow-up observations remain valid in such a scenario. For the targeted observations, the lower limits on the statistical spectral index lie in the range $\alpha_{\textrm{s}} > (-2.88, 0.55)$ for the sources listed in Table~\ref{tab:targeted_nondetections}. The targeted observations of FRB~20201124A yield $\alpha_{\textrm{s}} > 0.55$, making it not just the strongest constraint, but also the only FRB for which the repetition rate is confirmed to be lower at 150\,MHz relative to 600\,MHz. 

Assuming that emission from all FRB sources exhibits similar frequency dependence, we can constrain $\alpha_\textrm{s}$ for the ensemble spectrum of the FRB population. Since LOTAAS observations did not cover the entire localisation region for the majority of the sources in our search sample, we constrain the ensemble $\alpha_\textrm{s}$ probabilistically instead of directly solving Equation~\ref{eq:alpha_s}. 

\begin{figure}
\includegraphics[width=\columnwidth]{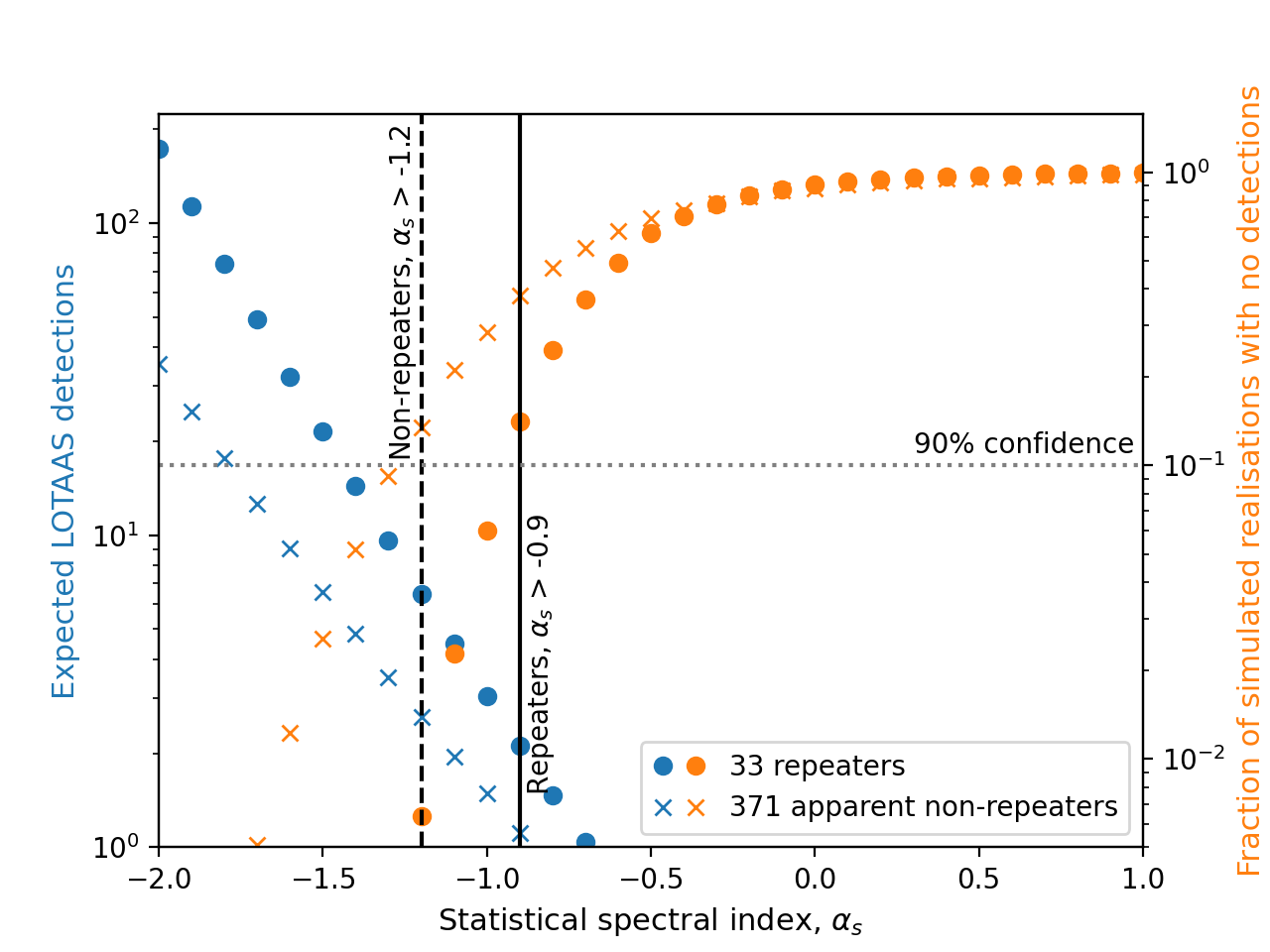}
    \caption{Constraints on the average statistical spectral index ($\alpha_\textrm{s}$) for 33 repeating FRB sources and 371 (apparent) non-repeating FRBs. The constraints are derived by simulating $10^4$ realisations of our LOTAAS search, with the average number of expected detections in these realisations plotted on the y-axis. Based on our non-detection, values of $\alpha_\textrm{s}$ less than the plotted 90\% confidence constraints can be ruled out as they predict LOTAAS FRB detections in $>90$\% of the simulated realisations.}
    \label{fig:indexoverall}
\end{figure}

We perform Monte Carlo simulations in which we iterate through trial statistical spectral indices (see Figure~\ref{fig:indexoverall}). For each trial index, we simulate 10,000 realisations of our search. In each realisation, we first sample $\lambda_\textrm{C}$ and $\gamma_\textrm{src}$ for each source from the distributions described earlier in this section. We then determine $\lambda_\textrm{L}$ by substituting the sampled values in Equation~\ref{eq:alpha_s}. Then we randomly sample the number of expected detections per source with LOTAAS from a Poisson distribution with a mean of $\lambda_\textrm{L}$. However, available LOTAAS beams only covered a fraction of the localisation region for several sources. For these sources, we set the expected detection count to zero for $x$\% of the realisations, where $x$ corresponds to the fraction of TABs overlapping with the localisation region for which data were not available. 
 
In summary, we simulate the total number of expected detections from 33 repeaters in LOTAAS data for 10,000 realisations of our search. The average detection count for these realisations is plotted in Figure~\ref{fig:indexoverall}. We use these detection counts to constrain the ensemble $\alpha_\textrm{s}$, ruling out values for which the detection count is non-zero in $>90$\% of the simulated realisations. For the repeaters in our sample, the resulting 90\% confidence constraint is $\alpha_\textrm{s} > -0.9$.  

Most of the constraining power comes from the highly active FRB~20201124A. A potential issue with using FRB~20201124A to determine $\alpha_\textrm{s}$ for repeaters is that its repetition rate is not representative of the FRB population in general. Only a few other repeaters, such as FRBs~20121102A, 20200120E, 20201130A, and 20220912A have comparable rates  \citep{li21,nimmo23,chime23,zhang23}. Another issue is that the sensitivity of LOFAR to FRB~20201124A is derived assuming a scattering timescale of 2\,$\mu$s at 600\,MHz, inferred from the scintillation bandwidth. If two distinct scattering screens exist in the direction of the source, as was the case for FRB~20110523 \citep{masui15}, the source could have another scattering timescale and potentially a higher fluence threshold. Therefore, we also report a constraint on the overall statistical spectral index by excluding FRB~20201124A. In this scenario, the constraint weakens to $\alpha_\textrm{s} > -1.6$.  

Assuming that the as-yet non-repeating sources will eventually repeat, we also constrain $\alpha_\textrm{s}$ for the ensemble spectrum of non-repeaters. We adopt the same approach as for the repeaters, sampling values of $\lambda_\textrm{C}$ within the
90\% confidence intervals derived by \citet{chime23}. We obtain a 90\% confidence constraint of $\alpha_\textrm{s} > -1.2$. This constraint is based on the non-detection of 371 apparent non-repeaters.  Although the localisation regions of 430 non-repeating sources overlapped with available LOTAAS TABs, we consider only those sources in our analysis for which a burst rate has been measured in the CHIME band.

We do not derive an ensemble constraint using the targeted follow-up observations. This is because these observations were triggered when the sources were active in the CHIME band, or a higher-frequency band, which biases the repetition rates in a way that cannot be adequately characterised.  

\subsection{Constraints on Source Environments}\label{sec:env}
A potential cause of the non-detection is a low-frequency spectral turnover due to absorption in the circumburst environment. Such a turnover is possible in the context of FRB source models, most of which involve young neutron stars which are likely to inhabit dense environments. Several absorption processes can be at play in such environments, such as stimulated Raman scattering, the Razin–Tsytovich effect, induced Compton scattering, and free-free absorption \citep{ravi19}. In this section, we focus on free-free absorption as it requires lower electron densities and temperatures than most other absorption processes. Exploring effects such as Razin–Tsytovich and induced Raman scattering will provide stronger constraints on the local environment \citep{ravi19}. By exclusively considering free-free absorption, we can report conservative constraints without making assumptions about emission properties, such as the brightness temperature for induced Compton scattering. 

We constrain the properties of the circumburst environments of FRBs, under the assumption that free-free absorption is rendering the low-frequency emission undetectable. We derive these constraints only for repeaters that exhibit high repetition rates and which were followed up with LOFAR within a few days of reported activity at higher frequencies (see Figure~\ref{fig:timeline_targeted_repeaters}). These criteria ensure that there was a reasonable chance of detecting a burst within the limited LOFAR observing time.

We evaluate whether the first criterion is met by scaling the CHIME repetition rate to the LOFAR fluence threshold using Equation~\ref{eq:alpha_s}. We solve this equation for 10,000 sets of $\gamma_\textrm{src}$ and $\lambda_\textrm{C}$, where each is sampled from the ranges listed in \S\ref{sec:alpha_s}. The median of the resulting distribution is assumed to be the expected LOFAR rate. Multiplying this rate by the duration of the targeted observations of each source provides the expected number of detections. If the latter is greater than 2.303, then the probability of detecting at least one burst is greater than 90\% as per Poissonian statistics (see, e.g., \citealt{gehrels86}). We use this threshold to determine whether there was a reasonable chance of detecting a burst from each source in our targeted observations. In solving Equation~\ref{eq:alpha_s}, we assume $\alpha_\textrm{s}$ = 0, i.e. any observed reduction in the rate at low frequencies is due to absorption and the intrinsic burst rate is independent of emission frequency. This is a conservative assumption as repetition rate has been observed to decrease with frequency for sources such as FRB~20121102A and FRB~20180916B \citep{houben19,chawla20,pastormarazuela21}.   

We constrain the properties of the local environment for the sources that meet both criteria, which in this case is only FRB~20201124A. We assume that the source is surrounded by a nebula of constant density. The emission at low frequencies would be attenuated, with the magnitude of attenuation depending on the electron density ($n_e$) and electron temperature ($T_e$) in the nebula. We simulate the attenuated fluence at a frequency $\nu$ as,
\begin{equation}\label{eq:spectra}
F_{\textrm{obs}, \nu} = F_{\nu_0} \bigg(\frac{\nu}{\nu_0}\bigg)^{\alpha} e^{-\tau_{\nu}},
\end{equation}
where the reference frequency $\nu_0 = 600 \ \textrm{MHz}$. In doing so, we assume the source emits a power-law spectrum with index $\alpha$. We note that $\alpha$ and $\alpha_\textrm{s}$ are distinct quantities in that $\alpha$ characterises the assumed power-law variation in burst intensity with frequency ($F \propto \nu^{\alpha}$). On the other hand, $\alpha_\textrm{s}$ characterises the frequency dependence of the repetition rate (at a given energy), which accounts for variation in both source activity and burst intensity with frequency.

The optical depth $\tau_\nu$ is defined as,
\begin{multline}
\tau_\nu = 3.014 \times 10^{-2} \ T_e^{-1.5} \ \bigg(\frac{\nu}{\textrm{GHz}}\bigg)^{-2} \biggl\{ ln \biggr[4.955 \times 10^{-2} \ \bigg(\frac{\nu}{\textrm{GHz}}\bigg)^{-1} \biggr] \\ + 1.5 \ ln (T_e) \biggl\} \ \bigg(\frac{\textrm{EM}}{\textrm{pc cm}^{-6}}\bigg)
\end{multline}
The above equation is derived by \citet{oster61} with the emission measure defined as $\textrm{EM} = n_e^2 R$. Here $R$ is the radius of the nebula for which we consider two values, 0.1 and 1 pc. For comparison, the Crab Nebula has a radius of $\sim1.7$ pc at an age of $\sim$970 years \citep{hester08}. The value for the reference fluence, $F_{\nu_0}$, is determined such that the simulated fluence averaged over 400--800\,MHz is equal to the mean fluence observed with CHIME for the FRB source under consideration. We set $\alpha$ equal to $-1.5$, i.e., the average spectral index determined by \citet{macquart19} through observations of 23 FRBs detected with the Australian SKA Pathfinder (ASKAP).    

\begin{figure}
\includegraphics[width=\columnwidth]{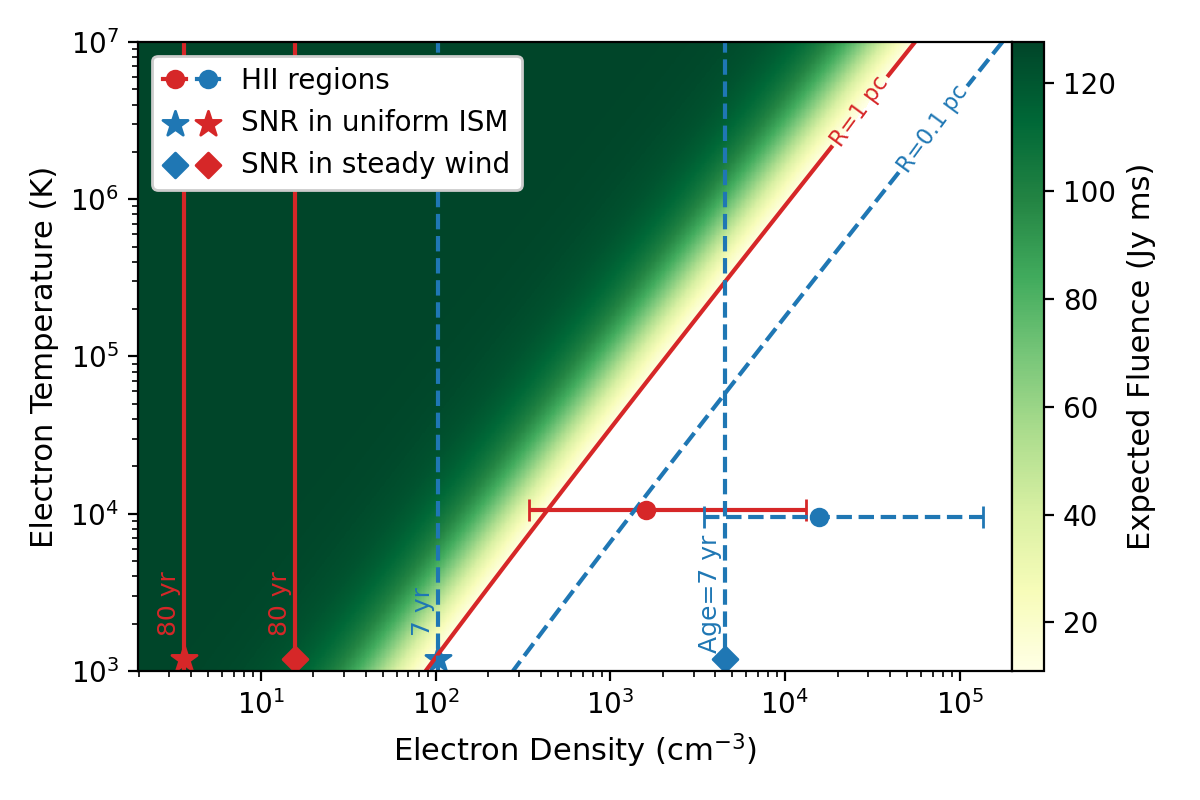}
    \caption{Constraints on the electron density and temperature of the local environment of FRB~20201124A derived based on the LOFAR non-detection of the source. The shaded region can be excluded as the local environment is expected to be optically thin to FRB emission at 150\,MHz in this part of the parameter space. The expected range of electron densities for H\textsc{ii} regions are derived based on the size-density relation determined by \citet{hunt09} and are coloured according to the radius. Although we adopt a temperature of 10$^4$ K for the H\textsc{ii} regions, the plotted data are slightly offset from the fiducial value for visual clarity. The expected electron densities and ages for supernova remnants evolving in a constant density ambient medium and into the stellar wind of the progenitor are also plotted \citep{piro18}, for the two fiducial radii values.}
    \label{fig:localenv}
\end{figure}

Integrating the resulting spectrum from 110 to 188\,MHz provides the expected fluence in the LOFAR band. The observed frequencies correspond to rest-frame frequencies of $(1+z)\, \nu$, where $z$ is the source redshift. We account for this redshift correction using the measured host galaxy redshift. The resulting fluence as a function of electron density and temperature is plotted in Figure~\ref{fig:localenv}.

The expected fluence for a certain $n_e$ and $T_e$ being greater than the LOFAR threshold implies that a nebula with those properties is optically thin to FRB emission in the LOFAR band. Therefore, if free-free absorption is the cause of the LOFAR non-detection, we can rule out a nebula with those properties as being associated with the source. This statement is valid for the shaded region in Figure~\ref{fig:localenv}.

We compare the resulting constraints with the electron densities and temperatures expected if FRBs originate from young neutron stars. In this scenario, FRBs could reside in supernova remnants or in H\textsc{ii} regions associated with recent star formation. The expected electron densities in H\textsc{ii} regions of the assumed radii (0.1 and 1 pc) are estimated using the size-density relation in \citet{hunt09} and plotted in Figure~\ref{fig:localenv}. The magnitude of absorption expected for these electron densities and typical temperatures of 10$^4$ K can easily reduce the intensity of low-frequency emission below the LOFAR threshold. Therefore, our non-detection is consistent with the source inhabiting a typical H\textsc{ii} region. 

For supernova remnants, we use the expected electron density and age estimated by \citet{piro18}. They determine the time evolution of the DM and radius for two scenarios, a supernova ejecta of 1 M$_\odot$ evolving into constant density ISM and into the stellar wind of the progenitor. In their estimates, the DM contribution from a central pulsar wind nebula is not included, though \citet{piro18} note it is likely small compared to that of the outer shocked material. We compute the electron density based on their DM and radius estimates. The resulting electron densities for supernova remnants with radii of 0.1 pc and 1 pc and ages of 7 yr and 80 yr, respectively, are plotted in Figure~\ref{fig:localenv}. On comparing these values with the constraints derived from LOFAR observations, we find that an extremely young supernova remnant (age $< 10$ yr) can easily render low-frequency emission undetectable with LOFAR. However, an older supernova remnant (age $\sim$ 100 yr) should be optically thin to low-frequency emission and is thus inconsistent with our non-detection.  

\section{Predictions for Future Searches}\label{sec:future}
No FRBs have been detected so far in untargeted searches at frequencies below 300\,MHz (see Table~1 in \citealt{sokolowski24} for a summary of the search parameters). The most sensitive of these searches was the LOFAR Pilot Pulsar Survey (LPPS) conducted by \citet{coenen14}, albeit with incoherent beam-forming and a low total observing time ($\sim$10 days). The LOFAR telescope is currently being upgraded, with LOFAR2.0 slated to have an increased computational capacity which will enable FRB searches over hundreds of tied-array beams.

Based on the measured FRB rate in the CHIME band \citep{chime21}, we predict the FRB detection rate for future searches with LOFAR2.0. The detection rate depends on several factors, including the redshift and energy distribution of FRBs, their scattering timescales, spectral indices and turnover frequencies (related to absorption processes in the local environment). We simulate these properties for $10^7$ FRBs to determine the fraction of sources for which the predicted fluence in the LOFAR2.0 band is greater than the expected system sensitivity.

We simulate the source redshifts, $z_\textrm{L}$,  assuming that their comoving number density follows the star-formation rate (see, e.g., \citealt{chawla22}). The assumption is motivated by models that invoke young neutron stars as FRB progenitors (see, e.g., \citealt{connor16, margalit18}). The assumption is also supported by a study of the observed redshift distribution of Murriyang (Parkes) and ASKAP FRBs by \citet{james22}, which suggests that the FRB population evolves with redshift in a manner consistent with, or faster than, the star formation rate. We only simulate redshifts up to a maximum of $z = 3$. We make this assumption since we use the CHIME/FRB rate as the reference all-sky rate and the maximum possible redshift inferred for the highest-DM event in the first CHIME/FRB catalogue is $z = 2.95$ \citep{chawla22}. 

We then simulate sky locations for each burst, in order to determine the Galactic contribution to burst DMs and scattering times. Simulating these locations also allows us to account for the varying sky temperatures while determining the LOFAR2.0 sensitivity to the simulated bursts. The sky locations are sampled uniformly over the entire northern sky, $\delta > 0^{\circ}$.

The DM of each FRB is simulated to include the contribution of different intervening media, 
\begin{equation}
\textrm{DM} = \textrm{DM}_\textrm{MW, ISM} +
\textrm{DM}_\textrm{MW, Halo} +
\textrm{DM}_\textrm{IGM} +\frac{\textrm{DM}_\textrm{Host + Local}}{(1+z)}.
\end{equation}
We calculate the Galactic DM, $\textrm{DM}_\textrm{MW, ISM}$, along the line-of-sight of each FRB using the NE2001 electron density model \citep{cordes02}. Several studies predict the DM of the Galactic Halo to be in the range, $10 < \textrm{DM}_\textrm{MW, Halo} < 100$ pc~cm$^{-3}$ \citep{dolag15,prochaska19,keating20}. We assume it to be 30 pc~cm$^{-3}$, following \citet{dolag15}. The IGM DMs are simulated as per the Macquart relation using the source redshifts based on the functional form in Equation~8 in \citet{zheng14}. The DM of the host galaxy including that of the local environment, $\textrm{DM}_\textrm{Host + Local}$ is drawn from a log-normal distribution with parameters set to those inferred using the first CHIME/FRB catalogue \citep{shin23}. The median of the sampled values is 84 pc~cm$^{-3}$ in the rest frame of the FRB and is reduced by $(1+z)$ to correct for the FRB redshift \citep{ioka03}. 

We then simulate the intrinsic width of each burst by sampling from the selection-corrected width distribution inferred for the first CHIME/FRB catalogue \citep{chime21}. Similarly, we sample scattering times from the corresponding CHIME/FRB selection-corrected distribution. We make several modifications to this simulated distribution. First, we truncate the distribution at 10 ms and only sample values below that threshold. We adopt this threshold as the CHIME/FRB rate is calculated for bursts with scattering times $< 10$ ms at 600\,MHz. Second, we scale the simulated scattering time distribution to 150\,MHz, assuming a power-law index of $-4$ for the frequency dependence \citep{bhat04}. The third modification to the simulated scattering times accounts for the difference in the observable redshifts for CHIME and LOFAR2.0. This is necessary as scattering timescales are implicitly dependent on source redshifts. The scattering contribution of material at the source redshift, i.e., in the host galaxy and circumburst environment, is reduced by a factor of $(1+ z)^3$ compared to that of the Milky Way \citep{macquart13}. 

In applying this correction, we make a simplifying assumption that negligible scattering arises in the IGM. This is justified by the predicted IGM scattering at 150\,MHz being $\sim$1 ms for a source with a DM$_\textrm{IGM}$ of 2,000 pc~cm$^{-3}$ and a turbulence injection scale associated with cosmic structure formation of 0.1 Mpc \citep{macquart13, zhu18}. This value is an order of magnitude lower than the sampling time for FRB searches with LOFAR2.0. The scattering arising in intervening galaxies (see, e.g., \citealt{vedantham19,faber24}) should also be corrected for the redshift it originates at. However, we do not apply that correction in this work as that would involve making assumptions about several parameters, such as the volume filling factor and the fraction of baryons in the CGM. 

For each simulated FRB, we determine the Galactic contribution to scattering using the NE2001 model \citep{cordes02} and initially subtract it from the simulated scattering time. Assuming the remaining scattering arises in the host galaxy and/or the circumburst environment, we scale the resulting scattering time as $(1 + z_\textrm{C})^3 / (1+ z_\textrm{L})^3$. The observable CHIME redshifts, $z_\textrm{C}$, are randomly sampled from the distribution inferred from CHIME/FRB observations by \citet{shin23}. The expected scattering time distribution for LOFAR2.0 is obtained by adding back the Galactic contribution after the redshift correction is applied. 

In these simulations, the intrinsic energy distribution of FRBs is modelled as a Schechter function \citep{schechter76}, for which the differential energy distribution has an exponential cut-off above a maximum energy $E_\textrm{max}$ and is defined as, 
\begin{equation}
\frac{dN}{dE} \propto E^{\gamma} \textrm{exp}\bigg(-\frac{E}{E_\textrm{max}}\bigg).
\end{equation}
The $\gamma_\textrm{src}$ in Equation~\ref{eq:energydist} is distinct from $\gamma$ in the above equation. The former is the power-law index of the energy distribution for an individual source and the latter is for the FRB population as a whole. We adopt a value of $-1.3$ for $\gamma$ and $2.4 \times 10^{41}$ erg for $E_\textrm{max}$ derived using CHIME/FRB data \citep{shin23}. There are other estimates of the maximum energy of FRBs in the literature \citep{luo20,james22,ryder23}. However, we adopt the CHIME/FRB value for consistency with the reference all-sky rate. Adopting any of the other estimates would not significantly change our results as the maximum energy estimated by \citet{shin23} is consistent within uncertainties with those derived in other studies.

The maximum energy estimate by \citet{shin23} assumes an intrinsic emission bandwidth of 1\,GHz. We adopt the same bandwidth for our simulated bursts and model their emission from 100\,MHz to 1.1\,GHz. The bursts are assumed to have a power-law spectrum with the index $\alpha$ being a variable parameter. We run the simulations thrice, each for a different value of $\alpha$ $(-0.5, -1.5$ and $-2.5)$. These values sample the full range of the constraints derived using both CHIME/FRB and ASKAP data. The measured value using CHIME/FRB data is $\alpha = -1.39^{+0.86}_{-1.19}$ \citep{shin23} and for ASKAP data is $\alpha = -1.5^{+0.2}_{-0.3}$ \citep{macquart19}.  

We also simulate a turnover in the spectrum, assuming that free-free absorption is at play in the circumburst environment. The turnover frequency for each FRB is determined using Equation~\ref{eq:spectra}, based on its EM and an electron temperature of 10,000 K (typical for H\textsc{ii} regions; \citealt{hunt09}). The EMs are sampled from a log-normal distribution, with a mean of $10^4$ pc cm$^{-6}$ and a standard deviation of 1.5 for the underlying normal distribution. These parameters allow sampling of EMs in the range $100-10^6$ cm$^{-6}$~pc, inferred through observations of GHz-peaked emission in pulsars \citep{rajwade16}. The constraint for the circumburst medium of FRB~20121102A, EM $> 8 \times 10^4$ cm$^{-6}$~pc \citep{cordes19}, also falls within the range of values that we simulate. To assess the dependence of rate predictions on the simulated EMs, we perform another iteration of the simulation with a higher mean of $10^5$ cm$^{-6}$~pc for the log-normal distribution.

The simulated spectrum for each FRB is calculated in the rest frame, and is corrected to the observed frame using the simulated redshifts. We then normalise each spectrum such that the energy emitted between 100\,MHz to 1.1\,GHz is equal to the simulated burst energy. The observed spectral energy densities are converted to fluence using the following equation,
\begin{equation}
F_{\nu} = \frac{E_{\nu} (1+z)^{2+\alpha}}{4 \pi D_L^2},
\end{equation}
where $D_L$ is the luminosity distance \citep{macquart18}. We assume that FRB searches with LOFAR2.0 will use all 24 Core stations with a bandwidth of 48\,MHz, centred at 150\,MHz. If the average fluence in this bandwidth is greater than the threshold computed using Equation~\ref{eq:threshold} for a $\textrm{S/N}_\textrm{th}$ of 7, then the simulated FRB is assumed to be detectable with LOFAR2.0. 

While calculating the fluence threshold, we set the gain to 6.8 K/Jy, i.e., four times that for the LOTAAS survey which uses only six stations. We apply no additional gain corrections with respect to elevation, thus assuming that the beam is pointed at zenith. The broadened pulse width in Equation~\ref{eq:threshold} includes a sampling time of 10 ms. The intra-channel dispersive smearing is calculated assuming that the data are coherently dedispersed every 50 pc cm$^{-3}$. This is the planned dedispersion strategy for the EuroFlash 24/7 commensal FRB search system which will be operational on LOFAR2.0 and ensures that the dispersive smearing at 150\,MHz is $<10$\,ms at all search DMs. 

\begin{figure}
\includegraphics[width=\columnwidth]{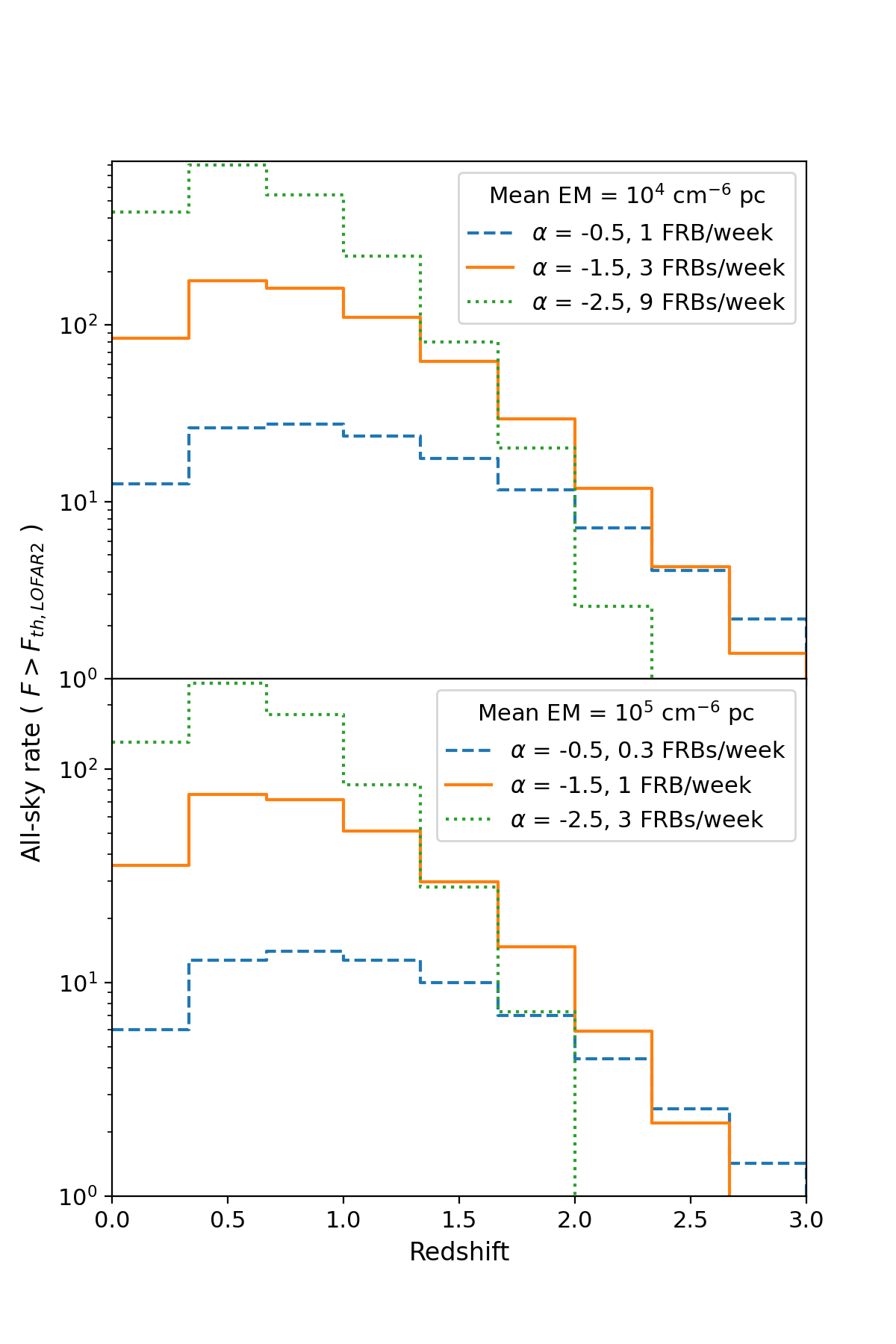}
    \caption{Predicted detection rate for an FRB search with the 24 LOFAR Core stations, an instantaneous field of view of 24 sq. deg. and an observing bandwidth of 48\,MHz. The rate is inferred by simulating a population of FRBs up to a maximum redshift of 3. The energy distribution of these FRBs is assumed to be a Schechter function with a power-law index of $-1.3$ and a maximum energy turnover of $2.4 \times 10^{41}$\,erg \citep{shin23}. The simulated FRBs are assumed to have a power-law spectrum with index $\alpha$ and a spectral turnover caused by free-free absorption. The turnover frequency for each FRB is determined based on its EM. These EMs are drawn from a log-normal distribution, the mean for which is indicated in the legend. The detection rate is normalised to the all-sky rate at 600\,MHz reported by \citet{chime21}.}
    \label{fig:rateprediction}
\end{figure}

The simulation thus returns the number of FRBs at $z < 3$ that are detectable with LOFAR2.0 out of the $10^7$ simulated events. The detection rate in the simulations is normalised to the all-sky rate at 600\,MHz of 525 FRBs per day above a fluence of 5 Jy ms reported by \citet{chime21}. The all-sky rate above the LOFAR2.0 sensitivity threshold is shown in Figure~\ref{fig:rateprediction} as a function of redshift, assumed mean EM, and spectral index of the FRB population. 

The simulations suggest that a week of observations with LOFAR2.0, with commensal FRB searching over a field of view of 24 sq. deg. could result in the detection of 0.3--9 FRBs per week. The predicted rates reduce by a factor of 2--3 if the mean assumed EM is varied from $10^4 \textrm{ to } 10^5 \textrm{ cm}^{-6} \textrm{ pc}$. This is expected since a higher EM implies a large optical depth, which can severely attenuate the low-frequency emission. 

As expected, the predicted rates are an order of magnitude higher if the spectral index is fairly steep, i.e. for an $\alpha = -2.5$ as compared to $\alpha = -0.5$. If the former index is more representative of the population, most low-frequency FRB detections will be at lower redshifts. However, if the index is fairly flat ($\alpha = -0.5$), then as many as 50\% of the detections will be between $1 < z < 3$. Discovery of these FRBs will be extremely useful for anchoring the Macquart relation at higher redshifts and in determining the associated uncertainties.
 
\section{Discussion}\label{sec:discussion}
In this work, we present results from the largest targeted FRB observing campaign at frequencies below 300\,MHz, in terms of the number of sources observed (14) and observing duration (a total of 252 hours). Similar campaigns in the past include an image-plane search with the MWA conducted by \citet{sokolowski18} for co-incident low-frequency (170--200\,MHz) counterparts of seven FRBs. These seven sources detected with ASKAP at 1.4\,GHz \citep{Macquart10} remained undetected in 3.5 hours of shadowed observations. \citet{tian2023} also performed a targeted search between 144--215\,MHz for bursts from five FRB sources in $\sim$24 hours of archival MWA data but did not detect any bursts.

Drawing on these prior campaigns, we adopted a few strategies to maximize our search sensitivity and chances of detection, namely:  
\begin{itemize}
    \item Following up an FRB source soon after detected activity at 600\,MHz or 1.4\,GHz. Most repeating FRBs exhibit non-Poissonian (clustered) bursting behaviour over longer timescales. Considering this, following up an FRB soon after reported activity at other frequencies maximized our chances of catching it in action. We note that our observations were not strictly simultaneous with CHIME/FRB detections. However, this should not significantly affect our chances of detecting bursts as repeater bursts are typically seen over a narrow bandwidth ($\sim 20\%$ fractional). The MWA non-detection of simultaneous low-frequency emission by \citet{sokolowski18}, non-detection of FRB~20121102A in multi-band observations \citep{houben19, Majid_2020_ApJL}, and non-simultaneous detections of FRB~20180916B \citep{chawla20,pastormarazuela21, pleunis21} in various frequency bands between 1.4\,GHz and 150\,MHz, provide evidence for the limited bandwidth of FRB emission and suggest that strict shadowing of higher frequency observations may not be necessary.
    \item Performing higher time resolution searches up to 82\,$\mu$s for sources with expected low-pulse broadening -- since many of the prolific repeating sources had reported burst widths and scattering measurements, we were able to optimize the time resolution of our searches to suit the expected range of widths at 150\,MHz.
    \item Performing subband searches with 15 overlapping subbands (with bandwidths as narrow as 12.5\% of the 78.8\,MHz bandwidth). This maximizes our sensitivity to the band-limited emission of repeating FRBs. Combined with the varying central frequency of burst emission, bursts can occupy as little as 25\% ($\sim20$\,MHz) of the $\sim$80\,MHz LOFAR bandwidth as seen in bursts from FRB~20180916B (the only published detections of an FRB below 300\,MHz \citealt{gopinath24}). We know of even lower fractional bandwidth occupation at higher frequencies, such as the 2\% fractional bandwidth (65\,MHz burst in 3.3\,GHz bandwidth) reported by \citet{Kumar21}. Our use of subband searches makes the reported fluence thresholds conservative, as they are based on a lower S/N obtained by averaging across the full band, rather than the higher S/N that would be achieved if a burst were confined to a sub-band.
    \item Zooming in on DM-space in the DM-time plots used for detection by the FETCH deep-learning classifier. Since a change in DM can reduce the S/N of the pulse more quickly at lower frequencies, this ensures that the `bow tie' structure in the DM-time plot has sufficient S/N to enable detection by the FETCH classifier. It also ensures better FETCH sensitivity to bursts with low fractional bandwidth.
\end{itemize}

We also adopted the final strategy outlined above in the search for repeat bursts from 463 CHIME FRBs in archival LOFAR observations. The non-detection of repeat bursts in both the targeted and archival LOFAR observations, despite the enhanced sensitivity and large observing time, results in strict constraints on the frequency dependence of FRB activity. 

\subsection{Constraints on frequency dependence of FRB activity}\label{sec:constraintsalpha}
We determine upper limits on the burst rates at LOFAR frequencies and then use these to constrain the frequency dependence of FRB activity. Our most constraining burst rate limit of $\lambda<0.05 \textrm{h}^{-1}$ (above a fluence threshold of 21 Jy ms) comes from observations of FRB~20200120E. For comparison, FRB~20180916B, which has the only published detections around 150\,MHz, exhibits a Poissonian burst rate of $0.08^{+0.01}_{-0.02} \textrm{h}^{-1}$, with 29 bursts detected in about 356 hours of exposure \citep{gopinath24}. This is not much larger than the strictest rate limit from our sample. However, in the case of FRB~20180916B, the burst activity is confined to only $\sim$25\% of a $\sim$16-day periodic window, meaning that a Poissonian rate model does not adequately describe its activity.

We compare our rate constraints with those presented by \citet{tian2023} who conducted targeted observations with the MWA around $185$\,MHz. They report burst rate limits of $<0.53 \textrm{h}^{-1}$ for FRB~20201124A (with 4 to 50 times lower sensitivity than LOFAR) and $<0.15 \textrm{h}^{-1}$ for FRB~20190117A (with sensitivity ranging between equivalent and 10 times lower than LOFAR). In contrast, we obtain a more stringent limit for FRB~20201124A of $<0.16 \textrm{h}^{-1}$ at similar frequencies, while our constraint for FRB~20190117A of $<0.21 \textrm{h}^{-1}$ is consistent with the MWA result.

Our rate limits provide lower limits on the statistical spectral index, ranging from $-2.88$ to $0.55$, for the 14 sources in our targeted observing campaign. Among these sources, the strongest constraint ($\alpha_s > 0.55$) is derived for FRB~20201124A, which implies that the repetition rate of this source decreases at lower frequencies. The statistical spectral index has previously been measured for very few sources which have been detected in multiple frequency bands. This includes the two repeating sources with known periodic activity, FRB~20121102A \citep{houben19, Gourdji2019} and FRB~20180916B \citep[$\alpha_{s, \textrm{0.35\,GHz}/\textrm{0.6\,GHz}}=-1.6^{+1.0}_{-0.6}$ and $\alpha_{s,\textrm{0.4\,GHz}/\textrm{0.8\,GHz}}=-0.6^{+1.8}_{-0.9}$, respectively]{chawla20, Sand2022}. 

The statistical spectral indices reported in the literature for FRB~20180916B suggest that it exhibits greater activity at lower frequencies and no spectral flattening or turnover. This is in stark contrast to the constraint we derive for FRB~20201124A. Our constraint is even more surprising, considering that negative spectral indices are expected as a consequence of FRB emission being coherent (see, e.g. \citealt{Kumar2017}). Additionally, giant pulses from pulsars, which, next to FRBs, are some of the most extreme coherent radio transients known, also exhibit negative spectral indices. \citet{Jankowski_2018} find a mean of $-1.6$, with potential variations due to frequency-dependent selection effects, for both normal and giant pulses from pulsars. More comparably, radio pulses from magnetars, on average, show flat ($\gtrsim -0.5$) spectral indices (see, e.g., \citealt{Camilo_2007_ApJ}). The spectral index and statistical spectral index are distinct properties in that the former describes the variation of the flux density of a source with frequency and the latter characterizes how the number of bursts emitted with a particular energy changes with frequency. Despite the distinction, a negative spectral index should, in general, result in a negative statistical spectral index. This is because brighter emission at low frequencies would enhance the likelihood of detection and lead to higher observed burst rates at those frequencies.

Similar to our constraint for FRB~20201124A, flat or positive statistical spectral indices have been inferred from observations of FRB~20121102A and FRB~20240619D. \citet{houben19} find that their limits on $\alpha_{s}$ for FRB~20121102A become flatter at lower frequencies, specifically, $\alpha_{s, \textrm{0.15\, GHz/1.2\,GHz}} > -0.5^{+0.2}_{-0.2}$, as compared to a much steeper measured $\alpha_{s}$ above 1\,GHz ($\alpha_{s, \textrm{1.4\,GHz/3\,GHz}} = -1.3^{+0.5}_{-0.6}$ and $\alpha_{s, \textrm{1.4\,GHz/2\,GHz}} = -2.5^{+1.2}_{-1.5}$). Their measurement is also consistent with the detection of only a single burst from FRB~20121102A with CHIME \citep{josephy19}, even though the source has been observed to be active for over 10 years thus far at frequencies above 1\,GHz. Since the expected scatter broadening for FRB~20121102A based on scintillation measurements is $\sim0.6$\,ms at 500\,MHz \citep{Ocker_2021}, the detection of this source at low frequencies should not be limited by temporal pulse broadening from scattering. Therefore, the aforementioned observations point towards a spectral turnover (either due to the emission mechanism or the circumburst environment; see \S\ref{sec:causes} below) as the cause of its non-detection.

Recently, \citet{Tian2025} discovered and monitored a repeating source FRB~20240619D with the MeerKAT telescope over a wide frequency range between 544\,MHz to 2.9\,GHz. They find that the source shows a spectral turnover, with a negative index at higher frequencies, $\alpha_{s, \textrm{1.3\,GHz/2.4\,GHz}} = -0.9^{+0.6}_{-0.6}$, and a markedly positive index at lower frequencies, $\alpha_{s, \textrm{0.8\,GHz/1.3\,GHz}} = 1.2^{+0.7}_{-0.6}$. Their result combined with the statistical spectral index constraints for FRB~20121102A and FRB~20201124A points towards the existence of a larger population of FRB sources that exhibit a turnover in burst rates at low frequencies.

We also use the  archival LOTAAS observations to derive a constraint on the statistical spectral index for the population as a whole (see Figure~\ref{fig:indexoverall}). The 90\% confidence constraints derived using LOTAAS observations are $\alpha_\textrm{s} > -0.9$ for the ensemble of 33 repeaters and $\alpha_\textrm{s} > -1.2$ for the 371 apparent non-repeaters. These limits can be compared with the constraints on the average spectral index for the FRB population derived using all-sky rates. The strongest of the existing constraints, $\alpha_\textrm{145\,MHz/1.4\,GHz} > 0.1$, was derived by comparing the upper limit on the all-sky rate at 145\,MHz with the 1.4-GHz detection rate \citep{karastergiou15}. However, this constraint assumes a pulse width of 5\,ms, which is an order of magnitude lower than the measured scattering timescale for the only FRB source detected at these frequencies ($\sim$50\,ms for FRB~20180916B; \citealt{pleunis21b, gopinath24}). \citet{chawla17} account for the reduced search sensitivity due to scattering and derive a constraint, $\alpha_\textrm{350\,MHz/1.4\,GHz} > -0.3$. In deriving these constraints, they assumed that the FRB population has a log-normal distribution of scattering timescales with a width equal to that determined for the Galactic scattering distribution.

Our population-level constraints, although consistent with those derived in the aforementioned works, are more robust for two reasons. Since more is known about FRB scattering as compared to when the earlier studies were published, we do not need to make any assumptions about the scattering time distribution. Instead, we use the measured scattering timescales of the sources in our sample while determining their search sensitivity. Second, we account for the reduction in sensitivity caused by the boxcars in our search pipeline recovering only a fraction of the fluence of highly scattered bursts (see \S\ref{sec:sensitivity}), which had not been accounted for in earlier studies.

Despite the robustness of our population-level constraints, they remain significantly weaker than the strongest per-source constraint for FRB~20201124A, primarily due to the limited observing time of only an hour per source for the LOTAAS observations. While our targeted observations had a larger exposure time per source, we exclude them from the population-level analysis due to the biases introduced by triggering these observations based on activity in the CHIME band. 

\subsection{What could explain our non-detections?}\label{sec:causes}
At low frequencies, we are biased to find lower burst rates due to high scattering times and sky temperatures from Galactic synchrotron emission. We account for both of these effects while determining the fluence thresholds. The CHIME/FRB scattering measurements used for the fluence threshold calculation are likely biased high due to intrinsically complex burst morphologies \citep{chime19}. As a result, the estimated scattering times at 150\,MHz for all sources, except FRB~20201124A, are likely to be upper limits. This implies that our quoted fluence thresholds are conservative. As for FRB~20201124A, the expected scatter broadening at 150\,MHz is negligible ($\sim$0.5\,ms). The LOFAR fluence threshhold of 9\,Jy~ms for FRB~20201124A shows that the array was particularly sensitive to this source compared to other sources in our sample, due to the source being close to zenith while transiting above the LOFAR core. Therefore, pulse broadening due to scattering is unlikely to have impacted its detectability.

As reduction in sensitivity due to increased scattering and sky temperature at low frequencies are already accounted for in our reported constraints, we investigate other causes of the non-detections below. For this discussion, we use FRB~20201124A as a guiding example to discuss possible reasons for the non-detections at low frequencies of the other sources in our sample and, more broadly, of the wider FRB population.

\subsubsection{Source activity}
The repetition of several highly active repeaters is shown to be non-Poissonian. Some repeaters, including FRBs~20200120E, 20200428D, and 20201124A, emit highly clustered bursts on timescales of seconds to hours \citep{nimmo23, kirsten21, lanman22}, often modelled as Weibull distributions \citep{Oppermann18}. This is somewhat apparent in the timeline in Figure~\ref{fig:timeline_targeted_repeaters}. FRB~20121102A and FRB~20180916B show periodic activity on much longer timescales of $\sim$160 and $\sim$16 days, respectively \citep{rajwade20,chime20b}.

For all sources, we minimized the risk of observing during times of low activity by scheduling the LOFAR targeted observations shortly after the source's detection with CHIME at 600\,MHz or a higher-frequency detection. Specifically for FRB~20201124A, the first set of LOFAR observations (4\,h, between MJDs~59480--59483) happened after CHIME detected it on MJD~59478 and during high levels of L-band activity \citep[for example]{Wu_2024}. The second set of LOFAR observations (10\,h, between MJDs~59613--59624) were triggered by a detection at 1.3\,GHz with the Westerbork radio telescope on MJD~59602 \citep{Ould-Boukattine_2022_ATel_a}. This set of observations happened during an active period of the source, spanning MJDs~59602--59640, at L-band and 2.2\,GHz  \citep{Kirsten_2024, Atri_2022_ATel, Bilous_2025, Takefuji_2022_ATel}. Although the source was not detected by CHIME at 600\,MHz during this epoch (this source transits CHIME for $\sim$3 min every day), it was found to be active in the 550--750\,MHz band at uGMRT \citep{Bilous_2025} on MJDs~59616 and 59617. We were observing the source with LOFAR on these days. For the rest of our source sample, most of the targeted observations were triggered within a few days of a CHIME detection of each source, as evidenced in Figure~\ref{fig:timeline_targeted_repeaters}.

It is indeed possible that the sources became inactive in the hours or days between the high-frequency detection and our triggered low-frequency follow-up, and that our observations therefore missed the window of enhanced activity. However, we account for this possibility by adopting extremely conservative 600-MHz rate measurements to derive the constraints on the statistical spectral index and the circumburst environment. We use the burst rates reported by \citet{chime23}, which are averaged over a three-year window. For most sources, the burst rates at 600\,MHz during our observations are likely to be much higher than these long-term averages, making our reported constraints quite conservative. Burst rates in times of activity can exceed the rates in inactive periods by orders of magnitude. For instance, FRB~20180916B has a burst rate of $0.17^{+0.28}_{-0.09} \textrm{h}^{-1}$  during its $\sim$4.3-day activity window at 150\,MHz, as compared to an upper limit of $0.0006 \ \textrm{h}^{-1}$ in the 12 days outside this window in its 16-day period \citep{gopinath24}.

Our population-level constraint using LOTAAS observations is also fairly robust to source activity in that it is based on non-detections of tens of sources and not all sources could be ``off'' at the time of our observations. Since this constraint, $\alpha_{s} > -0.9$, in addition to many of the per-source constraints are flatter than expected for coherent emission, we consider additional effects in the magnetosphere or the circumburst environment that could cause the decreased burst rate at low frequencies. 

\subsubsection{Emission mechanism}
There is growing evidence in support of a magnetospheric origin of FRBs (see, e.g., \citealt{Nimmo_2025}). Specifically in the case of FRB~20201124A, \citet{Niu2024} find bursts at L-band that show orthogonal jumps in their polarization angle, similar to single pulses from pulsars, further strengthening the argument for the magnetospheric origin of bursts from FRB~20201124A (coincidentally, two of three bursts that show this behaviour happened two days after the second LOFAR observing epoch of this source). 

A radius-to-frequency mapping effect in the source magnetosphere \citep{Lyutikov2020} has been conventionally accepted as an explanation for the often observed ``downward-drift'' in frequency seen across bursts from many repeating FRB sources \citep{hessels19}, including from FRB~20201124A (see \citealt{Kirsten_2024, Hilmarsson_2021b, Zhou_2022} for examples). In this model, the lower-frequency emission arises from larger emission radii. This is, in fact, the case for pulsar radio emission \citep{Manchester_1977_pulsars, Cordes_1978_ApJ, Phillips_1992_ApJ}. If one were to hypothesize that a relationship exists between burst energy and emission radius, due to, say, the magnetic field strength being lower at higher altitudes, or due to beaming geometry wherein the beaming angle is larger at larger altitudes leading to the energy being dissipated across a larger beaming angle --- then the non-detection of bursts at low frequencies can be attributed to a radius-frequency-energy mapping of the bursts. This can be tested by measuring energies of individual drifting components detected in bursts within the same (broad) band of the same telescope, in order to minimize complications in energy comparisons introduced when comparing across different instruments and frequency bands. 

There is also some evidence for lower-frequency bursts of the same source being larger in temporal width and fainter, independent of any effects introduced by scattering. Examples of increasing temporal width with reducing frequency can be seen often in bursts with multiple drifting components such as in \citet{Gajjar_2018} and \citet{Faber_2024b}. Increasing burst widths at lower frequencies might be related to the beaming angle at those frequencies. We know that such a dependency is used to explain similar behaviour observed for pulsar single pulses \citep{Radhakrishnan_1969}. If such a relationship exists for FRBs, then it could potentially explain the non-detection of bursts at low frequencies, since the detection metric is inversely dependent on the square root of the burst width.

\subsubsection{Absorption in local environment}
Among the sources we observed with LOFAR, FRB~20201124A had exposure during or shortly after an active phase and exhibited a sufficiently high repetition rate to allow for detection in the limited LOFAR observing time. Based on the CHIME/FRB rates, there was a $>$90\% probability of detecting a burst in the LOFAR observations. Consequently, we were able to place constraints on the properties of its circumburst environment (\S\ref{sec:env}).

To derive these constraints, we assumed a statistical spectral index of $\alpha_\textrm{s}$ = 0. As discussed in \S\ref{sec:constraintsalpha}, a negative value is more plausible based on the spectral indices observed for pulsars and the coherent nature of FRB emission, while a flatter spectral index is consistent with what is observed from magnetar radio pulses. Since we are already disentangling absorption (a phenomenon that can flatten the observed spectrum) from the statistical spectral index, our assumption of $\alpha_\textrm{s}$ = 0 is conservative. Considering the possibility that free-free absorption in the local environment of FRB~20201124A renders it undetectable at 150\,MHz, we assume the source is embedded within a nebula of constant density. The optical depth of such a nebula, which can be characterized in terms of its electron density and electron temperature, would then dictate burst detectability, given the fluence threshold for our search. We find that non-detection of the source at 150\,MHz is consistent with the source residing in a typical H\textsc{ii} region. If FRB~20201124A were instead to reside in a supernova remnant, we find that the age of the remnant must be very young ($\sim$10\,yr) to render it undetectable by LOFAR. A slightly older supernova remnant of age $\sim$100\,yr would not be optically thick enough to attenuate the low-frequency emission below the detection threshold of our search. 

A combination of factors --- short observing duration, and low sensitivity of our observations due to large scattering time, unreported CHIME/FRB fluence limits and width measurements --- prevent us from placing constraints on the properties of the local environments of the other FRBs in our sample. However, it is still possible that the other FRB sources occupy dense environments. We know of two FRB sources that are robustly associated with a persistent radio source (PRS) --- FRB~20121102A \citep{Marcote_2017}, and FRB~20190520B \citep{Niu_2022}. There are a few more tentative associations \citep{Ibik_2024, CHIME_FRBCollaboration_2025_arXiv,Bruni_2025}, including for FRB~20201124A \citep{Bruni_2024}. FRB~20121102 and FRB~20190520B are expected to reside in extreme, highly dynamic environments, given the large magnitude of ($>10^4$\,rad/m$^{-2}$) and drastic changes to the rotation measure (RM) observed in the bursts from these sources \citep{Hilmarsson_2021a,Anna-Thomas_2023}, in addition to scattering variability (independent of the RM variations) in FRB~20190520B \citep{Ocker_2023}. Such dynamic environments with potentially very high electron densities can easily render low frequency emission undetectable. Based on the scattering times reported in the CHIME/FRB catalog, \citet{chawla22} find that FRBs may inhabit more extreme environments than those of Galactic pulsars. In addition to having large line-of-sight electron density fluctuations that cause scattering, these environments may also have sufficiently high total electron densities to attenuate low-frequency emission.

\citet{Bruni_2024} find a relationship between the magnitude of RM and luminosities of the PRSs, thus proposing that PRSs are generated by a nebula in the FRB environment. It is worth noting that FRB~20121102A has an associated PRS, resides in an extreme environment and its bursts show a spectral turnover at frequencies below 400\,MHz \citep{houben19}. Similarly, our statistical spectral index limits for FRB~20201124A suggest a potential spectral turnover, and the source may also reside in an extreme environment associated with a PRS \citep{Bruni_2024}.

With powerful localization instruments such as the EVN and the CHIME outriggers, we are slated to have more unambiguous host galaxy associations and PRS associations. \citet{Law_2022} estimate that FRB sources with a PRS could constitute between $6-36$\% of the FRB population. Similar studies in the future on the extremely active repeating sources FRB~20220912A and FRB~20240114A, which have had significant LOFAR observing time as reported in this work, will be of interest. Furthermore, sources in environments currently opaque to low-frequency radio emission may in the future become transparent to such signals. \citet{Wang_2025_arXiv} recently show that the RM of FRB~20121102A has decreased by 70\% from $>10^{5}$\,rad/m$^{-2}$ \citep{michilli18b} to $\sim 31000$\,rad/m$^{-2}$ in a decade, along with a 7\% decrease in the local DM, hinting that the environment may be getting less extreme with time.

\subsection{Strategies for future low-frequency FRB follow-up}
In addition to placing constraints on FRB emission and environments, we predict FRB detection rates for untargeted searches with the upcoming LOFAR2.0 telescope. Our simulations, based on the CHIME/FRB detection rate and scattering time distribution, predict that LOFAR2.0 could detect between 0.3 and 9 FRBs per week. This wide range is due to the variation in the assumed intrinsic spectral indices of the simulated FRBs and the EMs of their circumburst environments. The assumed values of these parameters in the simulations are consistent with the non-detection of FRBs in surveys conducted at low frequencies. In particular, we also ran our simulations for the observing parameters of FRB surveys with LOFAR, MWA and GBT conducted by \citet{coenen14}, \citet{tingay15}, \citet{karastergiou15} and \citet{chawla17}. The simulations predicted zero FRB detections for any of these surveys across the full range of assumed spectral indices and EMs. The significant improvements in both sensitivity and observing time offered by LOFAR2.0 will make untargeted surveys an increasingly valuable strategy for searching FRBs and complement targeted searches of FRBs in constraining FRB emission mechanisms and the properties of their local environments. 

Results from our targeted searches provide several insights that can inform observing strategies for FRB searches with upcoming instruments such as LOFAR2.0 and the SKA-Low. Below, we summarize some key takeaways from this work:
    \begin{itemize}
        \item Targeted follow-up of repeating sources continues to be a valuable strategy to study the low-frequency emission of FRBs. As shown in this work, even in the event of a non-detection, meaningful constraints can be derived. Constraints from observations at even lower frequencies, such as those from the targeted FRB search campaign with the NenuFAR telescope at 10--85\,MHz \citep{Decoene_2023}, would further improve our understanding of the frequency dependence of FRB activity. 
    \item Future observations will have a greater chance of yielding detections or stronger constraints if active CHIME/FRB sources located at high Galactic latitudes and with low scattering times are picked as targets. An upcoming analysis (Gopinath et al., in prep.), which reports on the detection of FRB~20190212A at 150\,MHz, will explore what additional properties may make certain FRB sources more favourable candidates for low-frequency detection compared to those that remained undetected in this work.
    \item Statistical spectral index constraints can be useful in planning targeted follow-up observations. If tens of hours of observing time have been spent to follow up a particular FRB source at low frequencies without yielding any detections, then $\alpha_\textrm{s}$ should be evaluated by comparing with the burst rate at higher frequencies. If the constraint is already much stricter than that measured for known FRBs (see Figure~\ref{fig:indexpersrc}), then it is important to consider whether more observing time on the source is warranted and if the finite telescope time available can be spent on a different source. The code used in this work to compute the statistical spectral index and circumburst environment constraints has been made publicly available to facilitate future observational studies\footnote{\url{https://doi.org/10.5281/zenodo.16564505}}.
    \item Untargeted searches for one-off FRB sources and repeaters remain valuable as sources detected at low frequencies are likely to inhabit relatively clean environments with minimal scattering. Follow-up observations at higher frequencies could enable inclusion of these sources in samples used to constrain cosmological parameters such as the Hubble constant (see, e.g., \citealt{james22b}) and the IGM baryon fraction (see, e.g., \citealt{khrykin24}). \item As per our simulations in \S\ref{sec:future}, 
    there is potential to detect as many as $\sim$4 high redshift ($1<z<3)$ FRBs in a week of observations with LOFAR2.0. High-redshift FRBs might be more easily detectable at low frequencies due to their emission being redshifted. These FRBs can enable more precise constraints on cosmological parameters than their low-redshift counterparts since the uncertainties in the DM of the intergalactic medium greatly reduce with redshift \citep{zhang18}.
    \end{itemize} 

\section{Conclusions}\label{sec:conclusions}

In this paper, we present results from targeted LOFAR observations of CHIME-discovered repeaters --- the largest such campaign below 300\,MHz, with 252 observing hours across 14 sources. We also search for repeat bursts in archival LOFAR observations, from the LOTAAS survey, at the locations of CHIME repeaters and apparent non-repeaters. This provided another 473\,hrs of exposure to CHIME-discovered FRBs.

\begin{itemize}
    \item We detect no bursts at 150\,MHz from the 14 targeted repeating FRB sources, with total exposures ranging between 3 to 60 hours per source. 
    \item We detect no repeat bursts from 33 CHIME/FRB repeating FRB sources, 10 candidate repeaters and 430 apparently non-repeating sources in archival LOTAAS data, with 1\,h exposure time on each source.
    \item The non-detections result in a population-level constraint of $\alpha_{s, 135\,\rm{MHz}/600\,\rm{MHz}} > -0.9$ for repeaters. The statistical spectral index ($\alpha_{s}$) characterises the power-law scaling of fluence-normalised activity rates in different frequency bands, and thus parametrises the variation with frequency of both source activity and burst energy.
    \item Our strongest per-source constraint of $\alpha_{s} > 0.55$ from FRB~20201124A suggests decreasing burst rates at lower frequencies, which is contrary to the negative index expected due to the coherent nature of FRB emission, but consistent with the flat spectra observed from magnetar radio bursts.
    \item We argue against low source activity as being the cause of these nearly flat statistical spectral indices, although we cannot definitively rule out the possibility of reduced activity during our observations. We propose radius-to-frequency/energy mapping effects in the magnetosphere that could potentially explain the non-detection of low-frequency emission from the sources in our sample. Another likely explanation is that a significant fraction of FRBs inhabit dense environments which absorb or scatter low-frequency emission. 
    \item Non-detection of low-frequency emission from these repeating FRB sources enables us to constrain the properties of their local environments. If free-free absorption is the reason behind the non-detection of FRB~20201124A, we conclude that the source must inhabit a high-electron-density environment such as an H\textsc{ii} region or a very young ($\sim10$\,yr old) supernova remnant. A similar explanation could hold true for other sources in our sample. However, our observations lacked the necessary observing time and/or sensitivity to determine this conclusively.
    \item Our findings have implications for future low-frequency FRB follow-up. Targeted observations remain effective for detecting known sources that are active at low radio frequencies. However, untargeted low-frequency surveys are also valuable, as they may uncover sources with steep spectral indices that are missed at higher frequencies but reside in clean environments. Such low-frequency detections are particularly well-suited for cosmological applications, given the reduced environmental contamination. Our simulations predict that commensal FRB searching at 150\,MHz with LOFAR2.0 could result in the detection of 0.3--9 FRBs per week, with an instantaneous field-of-view of 24 sq. deg. using the 24 LOFAR Core stations. 

\end{itemize}

\section*{Acknowledgements}\label{sec:acknowledgements}
The AstroFlash research group at McGill University, University of Amsterdam, ASTRON, and JIVE is supported by: a Canada Excellence Research Chair in Transient Astrophysics (CERC-2022-00009); the European Research Council (ERC) under the European Union’s Horizon 2020 research and innovation programme (‘EuroFlash’; Grant agreement No. 101098079); and an NWO-Vici grant (‘AstroFlash’; VI.C.192.045). This paper is based (in part) on data obtained with the LOFAR telescope (LOFAR-ERIC) under project codes LT2\_003, LC3\_014, LC4\_030, LT5\_004, LC9\_023, LT10\_005, LT16\_008, LC17\_011, DDT18\_002 and LC20\_041. LOFAR \citep{vanhaarlem13} is the Low Frequency Array designed and constructed by ASTRON. It has observing, data processing, and data storage facilities in several countries, that are owned by various parties (each with their own funding sources), and that are collectively operated by the LOFAR European Research Infrastructure Consortium (LOFAR-ERIC) under a joint scientific policy. The LOFAR-ERIC resources have benefited from the following recent major funding sources: CNRS-INSU, Observatoire de Paris and Université d'Orléans, France; BMBF, MIWF-NRW, MPG, Germany; Science Foundation Ireland (SFI), Department of Business, Enterprise and Innovation (DBEI), Ireland; NWO, The Netherlands; The Science and Technology Facilities Council, UK; Ministry of Science and Higher Education, Poland. This publication is part of the project LOFAR Data Valorization (LDV) [project numbers 2020.031, 2022.033, and 2024.047] of the research programme Computing Time on National Computer Facilities using SPIDER that is (co-)funded by the Dutch Research Council (NWO), hosted by SURF through the call for proposals of Computing Time on National Computer Facilities. This work used the Dutch national e-infrastructure with the support of the SURF Cooperative using grant no. EINF-3440. We thank SURF (\url{www.surf.nl}) for the support in using the National Supercomputer Snellius. We also thank the operators of LOFAR for performing the LOTAAS observations. P.C. would like to thank Chia Min Tan for providing an overview of the LOTAAS processing pipeline at the beginning of this project. Z.P. is supported by an NWO Veni fellowship (VI.Veni.222.295). D.M. acknowledges support from the French government under the France 2030 investment plan, as part of the Initiative d'Excellence d'Aix-Marseille Universit\'e -- A*MIDEX (AMX-23-CEI-088). We acknowledge use of the CHIME/FRB Public Database, provided at \url{https://www.chime-frb.ca/} by the CHIME/FRB Collaboration, and the CHIME/FRB VOEvent Service.
\section*{Data Availability}
The LOFAR observations are available on the LOFAR Long Term Archive under project codes LT2\_003, LC3\_014, LC4\_030, LT5\_004, LC9\_023, LT10\_005, LT16\_008, LC17\_011, DDT18\_002 and LC20\_041. The relevant code and data products for this work are available on Zenodo (DOI:
10.5281/zenodo.16564505).

\bibliographystyle{mnras}
\bibliography{lotaas} 

\bsp	
\label{lastpage}
\end{document}